\documentclass[]{aa}  
\usepackage{amsmath}
\usepackage[normalem]{ulem}

\usepackage[dvipsnames]{xcolor}
\usepackage{float}
\usepackage{graphicx}
\usepackage{gensymb}

\usepackage{lscape}

\usepackage{graphicx}
\usepackage{multicol, blindtext}
\usepackage{breqn}
\usepackage{txfonts}
\usepackage[colorlinks=true,urlcolor=blue,citecolor=blue,linkcolor=blue,breaklinks=true]{hyperref}
\begin{document}

    \title{Spatio-temporal analysis of chromospheric heating in a plage region}
    %\subtitle{A spatio-temporal analysis of radiative losses from high-spatial resolution observations}

   \author{%R.~Morosin
          %\and
          %J.~de la Cruz Rodr\'iguez
          %\and
          %G.~J.~M.~Vissers
          R. Morosin
          \and
          J. de la Cruz Rodr\'iguez
          \and
          C. J. D\'iaz Baso
          %\and
          %C. Robustini
          \and
          J. Leenaarts
          }

   \institute{Institute for Solar Physics, Dept. of Astronomy, Stockholm University,               AlbaNova University Centre, SE-10691 Stockholm, Sweden\\
                \email{roberta.morosin@astro.su.se}
             }

   \date{Received XXX; accepted XXX}

% \abstract{}{}{}{}{} 
% 5 {} token are mandatory
 
  \abstract
 %{The energy balance of the solar chromosphere in active regions has been greatly discuss by the solar community. Radiative losses are predicted by several studies, but other types of mechanisms need to provide the energy necessary to sustain the chromosphere. Magnetic field related processes and different types of waves are the main proposed candidates.}
 {Our knowledge of the heating mechanisms that are at work in the chromosphere of plage regions remains highly unconstrained from observational studies. While many heating candidates have been proposed in theoretical studies, the exact contribution from each of them is still unknown. The problem is rather difficult because there is not direct way of estimating the heating terms from chromospheric observations.}
 {The purpose of our study is to estimate the chromospheric heating terms from a multi-line high spatial-resolution plage dataset, characterize their spatio-temporal distribution and set constraints on the heating processes that are at work in the chromosphere.}
 %{In order to obtain the model atmosphere from the spectropolarimetric observations, we inverted the first time step with the STockholm inversion Code (STiC). Instead of inverting every time step with the inversion code, we wrote a fully-connected neural network (FCNN) that, after the training, was able to reproduce the model atmosphere from the observational data. The radiative losses for 21 time steps ($\sim12$min) were obtained using STiC in synthesis mode, starting from the model atmospheres. }
 {We make use of non-local thermodynamical equilibrium (NLTE) inversions in order to infer a model of the photosphere and chromosphere of a plage dataset acquired with the Swedish 1-m Solar Telescope (SST). We use this model atmosphere to calculate the chromospheric radiative losses from the main chromospheric cooler from \ion{H}{i}, \ion{Ca}{ii} and \ion{Mg}{ii} atoms. In this study, we approximate the chromospheric heating terms by the net radiative losses predicted by the inverted model. In order to make the analysis of time-series over a large field-of-view (FOV) computationally tractable, we make use of a neural network which is trained from the inverted models of two non-consecutive time-steps. We have divided the chromosphere in three regions (lower, middle, upper) and analyzed how the distribution of the radiative losses is correlated with the physical parameters of the model.}
 %{We have obtained the radiative losses for the active region for the first time step of the observations. They are calculated over three integration areas, corresponding to lower, middle and upper chromosphere, employing bound-bound and bound-free transitions of \ion{Ca}{ii}, \ion{Mg}{ii} and \ion{H}\ atoms. The radiative losses are well correlated with the temperature structure and are confined in between pores and magnetic structures. As soon as the canopy is formed in the upper chromosphere, they expand covering the active region. \ion{Ca}{ii} is the main contributor in the lower chromosphere, while \ion{H}\ and \ion{Mg}{ii} dominate in the upper layers. Our results of the time series analysis suggest that there is a sinusoidal behaviour in the evolution of both the radiative losses and $v_{LOS}$ with time. The average amplitudes inside the plage region are $\overline{amp}_{Q}=5.2$ kW m$^{-2}$ and $\overline{amp}_{v_{LOS}}=2.8$ km/s, while the average periods are $\overline{p}_{Q}=5.2$ min and $\overline{p}_{v_{LOS}}=5.5$ min.}
 {In the lower chromosphere, the contribution from the \ion{Ca}{ii} lines is dominant and predominantly located in the surroundings of the photospheric footpoints. In the upper chromosphere, the \ion{H}{i} contribution is dominant. Radiative losses in the upper chromosphere form a relatively homogeneous patch that covers the entire plage region. The \ion{Mg}{ii} also peaks in the upper chromosphere. Our time analysis shows that in all pixels, the net radiative losses can be split in a periodic component with an average amplitude of $\overline{amp}_{Q}=7.6$~kW~m$^{-2}$ and a static (or very slowly evolving) component with a mean value of -26.1~kW~m$^{-2}$. The period of the modulation present in the net radiative losses matches that of the line-of-sight velocity of the model.}
 {Our interpretation is that in the lower chromosphere, the radiative losses are tracing the sharp lower edge of the hot magnetic canopy that is formed above the photosphere, where the electric current is expected to be large. Therefore Ohmic current dissipation could explain the observed distribution. In the upper chromosphere, both the magnetic field and the distribution of net radiative losses are room-filling and relatively smooth, whereas the amplitude of the periodic component is largest. Our results suggest that acoustic wave heating may be responsible for one third of the energy deposition in the upper chromosphere, whereas other heating mechanisms must be responsible for the rest: turbulent Alfv\'en wave dissipation or ambipolar difussion could be among them. Given the smooth nature of the magnetic field in the upper chromosphere, we are inclined to rule out Ohmic dissipation of current sheets in the upper chromosphere. }
   \keywords{ polarization -- Sun: chromosphere  -- Sun: faculae, plages}

   \maketitle
%
%-------------------------------------------------------------------

\section{Introduction}\label{sec:intro}
The heating of the solar chromosphere and corona remains one of the foremost questions in solar and stellar physics. The chromosphere is on average radiating 4~kW m$^{-2}$ in the quiet Sun and 20~kW m$^{-2}$ in active regions \citep{1981ApJS...45..635V, 1977ARA&A..15..363W}. At least, that energy must be transported and deposited into the chromosphere at any time by heating mechanisms. Although we cannot measure chromospheric heating terms directly, we can estimate them by assuming that they equal the radiative losses in the main chromospheric coolers, typically strong chromospheric lines and continua from the \ion{H}{i}, \ion{Ca}{ii} and \ion{Mg}{ii} atoms.

The physics and heating of plage regions have puzzled our community since the 70's. Three recent independent studies have attempted to infer the strength and stratification of the magnetic field in plage targets \citep[][]{2020A&A...642A.210M,2020A&A...644A..43P,2021SciA....7.8406I}. They found amplitudes of approximately $|B_\parallel|\sim300 - 400$~G, depending on the spectral line and target under analysis. In particular \cite{2020A&A...642A.210M} reconstructed the canopy effect of the magnetic field in the chromosphere. The magnetic field is very concentrated in the intergranular lanes in the photosphere and it expands horizontally as we move up in the atmosphere, forming a hot magnetic canopy over the photosphere. Because of the sharp lower boundary of the canopy, the authors speculated that current sheets should be present in this boundary, purely from the application of $\boldsymbol{j} = \nabla \times \boldsymbol{B}/\mu$. Those currents could lead to Ohmic dissipation at the lower boundary of plage, causing heating at the base of the chromosphere. 

Furthermore, modelling chromospheric lines from plage observations have typically required very large values of microturbulence, typically up to 10 km~s$^{-1}$ \citep[e.g.,][]{1974SoPh...39...49S,2015ApJ...809L..30C, 2015ApJ...799L..12D,2019ARA&A..57..189C}. A more recent study using ALMA and IRIS observations have further constrained this value to an average value of 5~km~s$^{-1}$ \citep{2020A&A...634A..56D}. Whether these large values of microturbulence are related to turbulent velocity fields, sharp gradients along the line-of-sight induced by a hot magnetic canopy above the photosphere \citep{1994ApJ...424.1014S,2013ApJ...764L..11D,2015A&A...576A..27B,2020A&A...642A.210M}, answering this question is entangled with the enhanced values of radiative losses that have been reported in plage in comparison with the quiet Sun.

In the chromosphere magnetic forces equal those from pressure gradients, leading to a complex and very dynamic force balance. Therefore magnetoacoustic waves, turbulent Alv\'en wave dissipation, magnetic reconnection, Ohmic current dissipation, viscous heating and ambipolar diffusion can all contribute to the heating of the chromosphere \citep[see,][and references therein]{2008ApJ...680.1542H,2011ApJ...736....3V,2012ApJ...747...87K,2014masu.book.....P,2017Sci...356.1269M,2018ApJ...862L..24P,2019ApJ...879...57B,2020ApJ...894L..17Y,2021A&A...647A.188D,2022arXiv220203955D}. The exact set of processes that are at work in plage and their contribution to the energy budget remains highly unconstrained from observations. A great discussion and exhaustive review about the potential contribution of different heating mechanisms in plage is presented by \citet{2021ApJ...921...39A}. They also analyzed the correlations of the radiative flux in the \ion{Mg}{ii}~h\&k lines and with the magnetic field strength that was inferred from observations in the \ion{He}{i}~10830~\AA\ line. Their conclusion was that Alfv\'en waves or ion-neutral collisions could be heating plage regions. They could not find a clear correlation with the electric current in their results.

Inversion methods allow for the inference of a model atmosphere of the photosphere and chromosphere by iteratively modifying the physical parameters of a model atmosphere to reproduce the observed full-Stokes spectra. They can be used for a variety of spectral lines and it is also possible to include non local thermodynamical equilibrium (NLTE) effects \citep{2008ApJ...683..542A,2015A&A...577A...7S, 2018A&A...617A..24M, 2019A&A...623A..74D,2022arXiv220202226R}. NLTE inversions are computationally expensive and time consuming, so studying the evolution of the atmosphere parameters of an entire time series observation could become prohibitive. The great advantage of inversions is that we can use the inferred model atmosphere to calculate radiative losses in the chromosphere \citep[see e.g.,][]{2020A&A...642A..52A,2021A&A...647A.188D} and thereby, obtain a lower-limit estimate of the chromospheric heating terms. To our knowledge, there is no other way to estimate radiative losses from observational data.

In this study, we have made use of a subset of a long time series to calculate plage models from NLTE inversions. We have used the resulting model atmospheres to train a neural network that can quickly predict the model atmosphere for the rest of the dataset, in a similar way to \citet{2019A&A...626A.102A} or  \citet{2020A&A...637A...1K}. The underlying assumption for this approach to work is that the training set is statistically representative of the entire time series. The quantities included in a model atmosphere are the gas temperature $T$, the line-of-sight velocity $v_{LOS}$, the turbulence velocity $v_{turb}$, and the parallel and perpendicular components of the magnetic field, respectively $B_{||}$ and $B_{\perp}$. 

We have calculated the net chromospheric radiative losses for all time steps of the series in order to better understand the distribution of radiative losses in a plage target, as well as the time evolution. In our analysis, we study some correlations with other physical parameters in order to suggest what heating mechanisms could be at work.

\section{Observations and data reduction}
 The target of interest is the NOAA2591 and it was observed with the Swedish 1-m Solar Telescope \citep[SST;][]{scharmer2003} on the 14th September 2016 at 08:26 UT. It is a plage region located at ($X$,$Y$) = (424\arcsec,-16\arcsec), that corresponds to a viewing angle of $\mu=0.90$. The CRisp Imaging Spectro-Polarimeter \citep[CRISP;][]{scharmer2008} and the CHROMospheric Imaging Spectrometer \citep[CHROMIS;][]{2017psio.confE..85S} were used in order to obtain observations in \ion{Ca}{ii}~8542~\AA (full-Stokes), \ion{Fe}{i}~6302~\AA\ (full-Stokes) and \ion{Ca}{ii}~H\&K (intensity only).
 
 \ion{Ca}{ii}~8542~\AA\ was sampled at 21 wavelength positions. The distance between each position was $\Delta \lambda = 85$\,m\AA, except for two points in the far wings of the line located at $\pm 1700$\, m\AA\ from line center. \ion{Fe}{i}~6302~\AA\ was recorded at 16 wavelength positions, of which, one at line center and two of them located in the red part part of the line at $\Delta \lambda = 40, 80$\,m\AA. The other 13 points are located at $\Delta \lambda = -40$\, m\AA\ from line center. The data detected with CRisp have a cadence of $\Delta t =37$s. In total the observation with the instrument lasts for $t\sim22$min.
 
\begin{figure*}[!ht]
        \centering
        \includegraphics[width=0.87\textwidth, trim=0.3cm 0.2cm 0.2cm 0.2cm, clip]{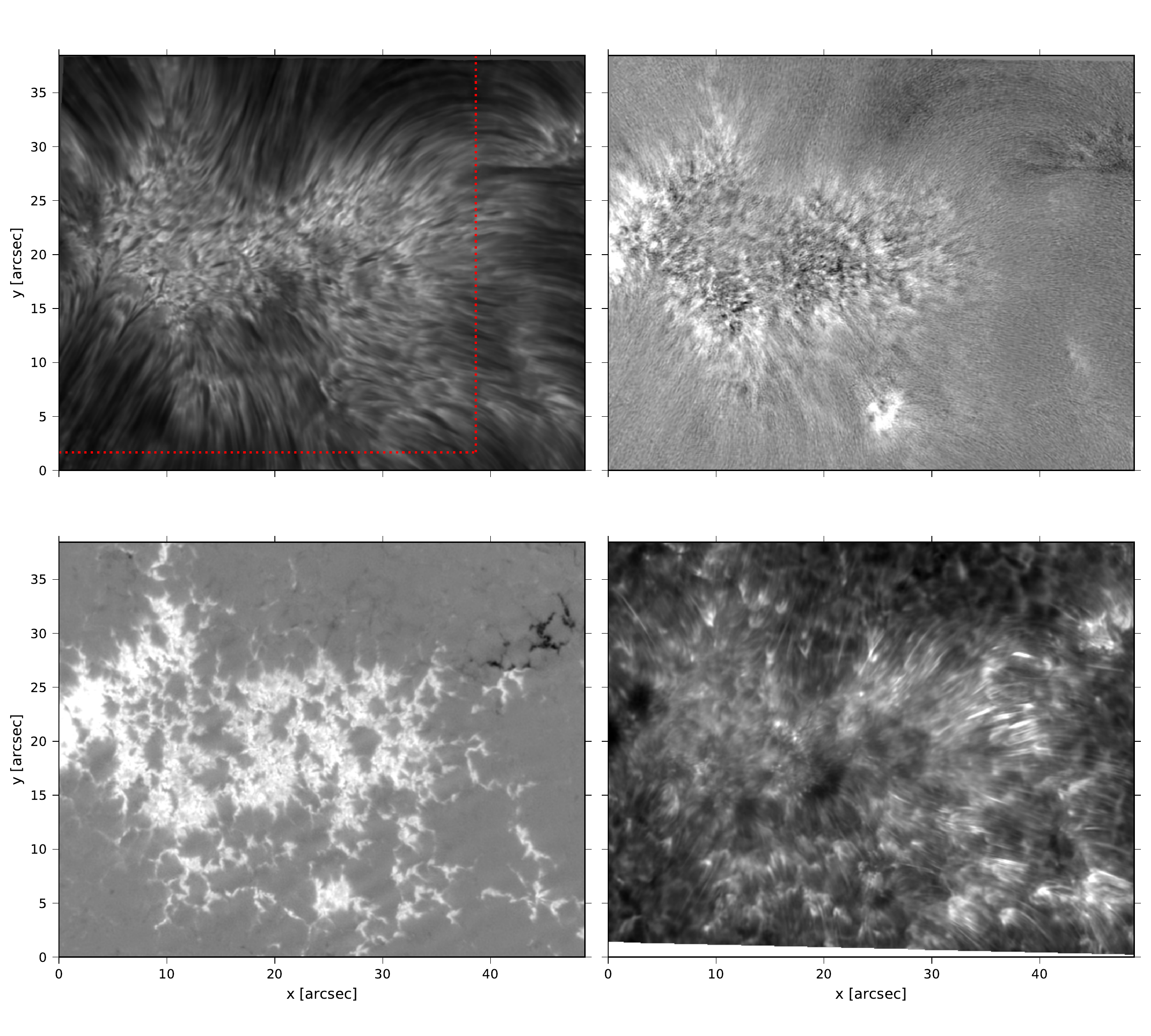}
        \caption{Overview of the observation in \ion{Ca}{ii}~8542~\AA, \ion{Fe}{i}~6302~\AA\ and \ion{Ca}{ii} {K}. \emph{Top row:} From the left Stokes $I$ at $\Delta \lambda=$85 m\AA\ and $V/I$ at $\Delta \lambda=$170 m\AA\ from line center. \emph{Bottom row:} From the left Stokes $V/I$ of \ion{Fe}{i}~6302~\AA\ at $\Delta \lambda=$40 m\AA\ from line center and Stokes $I$ of \ion{Ca}{ii} {K} at $\Delta \lambda=$470 m\AA. The red box indicates the area used in our study.}
        \label{stokes_data_ca}
\end{figure*}
 
The \ion{Ca}{ii}~H\&K lines were sampled at 21 wavelength positions, of which 19 at a distance of $\Delta \lambda=$79 m\AA\ from each other, and the two extra-wing points at a distance of $\Delta \lambda=$1.25 \AA from line center. A continuum position at 4000 \AA was also recorded. CHROMIS has a cadence of $\Delta t =16$s. In order to match the observations with the two instruments, for each snapshot of CRisp, we chose the closest snapshot of CHROMIS in time.
 
After the acquisition, the data have been reduced using the SSTRED pipeline \citep{jaime2015, 2021A&A...653A..68L}. In order to take into account atmospheric effects, the data have been also processed using the Multi-Object Multi-Frame Blind Deconvolution method (MOMFBD) described in \citet{vannoort}. Then the dataset has been properly aligned since the two instruments have a different pixel scale. We also have used the python package ISPy to handle the data and metadata of the data cubes \citep{2021zndo...5608441D}.
 
An overview of the observed active region is presented in Fig.~\ref{stokes_data_ca}. The panels depict the observations in \ion{Ca}{ii}~8542~\AA, \ion{Fe}{i}~6302~\AA\ and \ion{Ca}{ii} {K}. Stokes $I$ and $V/I$ of \ion{Ca}{ii}~8542~\AA\ are shown at $\Delta \lambda=$85 m\AA\ from line center. The upper left panel of Fig.~\ref{stokes_data_ca} illustrates the plage region in the chromosphere, with typical features as fibrils, extending from the center of the region towards the outside. There are two different polarity patches in the field-of-view (FOV). The \ion{Ca}{ii}~K wing image in the lower-right panel shows low-lying bright structures connecting both polarities.
 
 %Stokes $V/I$ of \ion{Fe}{i}~6302~\AA\ (bottom left panel) confirms the presence of an opposite polarity in the upper right corner of the FOV. Stokes $I$ of \ion{Ca}{ii} {K}\ (bottom right panel) is represented at the wings of the line: also in this line, bright fibrils directed outwards the active region are detected.

\section{Data analysis}
\subsection{Inversions} 

To estimate the thermodynamic and magnetic properties of the region we have performed non-LTE inversions, using the STockholm inversion Code (STiC) \citep{2016ApJ...830L..30D, 2019A&A...623A..74D}. It is a modified version of the radiative-transfer RH code \citep{2001ApJ...557..389U} and it includes a fast approximation to calculate the effects of partial redistribution \citep[PRD, see][]{2012A&A...543A.109L}. The inversion engine of STiC includes an equation of state extracted from the Spectroscopy Made Easy \citep[SME,][]{2017A&A...597A..16P}. The radiative transport equation is solved using a cubic Bezier solver \citep{2013ApJ...764...33D} of the polarized transfer equation.

%In order to minimize the difference between the observed and synthetic spectra and to obtain the best guess for the model atmosphere parameters, the Levenberg-Marquardt algorithm is applied. This algorithm iteratively compares the spectra and calculate the $\chi^2$. The algorithm makes the solution converge when a minimum $\chi^2$ is reached. STiC has been intensively used in many targets, from plage regions \citep{2020A&A...644A..43P} to different types of solar flares \citep{2021A&A...645A...1V, 2021A&A...649A.106Y}. 

%For our analysis we decided to treat the \ion{Ca}{ii} atom in non-LTE, while the \ion{Fe}{i} atom in LTE conditions. In particular for \ion{Ca}{ii}~{K} line, partial redistribution effects has been taken into account. Additionally, we also chose to perform the inversion in column mass ($cmass$), instead of $\log(\tau)$, the optical depth scale, that is most commonly used. Working in column mass will simplify the calculation of the z-scale in Section~\ref{sec:radiative} and also will expand the region of our interest, allowing to put more nodes and have a better sampling of the chromosphere \citep{2019A&A...623A..74D}.
The spectra in the \ion{Ca}{ii} spectral lines were calculated in NLTE by assuming statistical equilibrium and plane-parallel geometry. Furthermore, PRD effects were explicitly included in the calculations of the H\&K lines. The \ion{Fe}{i} lines were synthesized assuming local thermodynamical equilibrium. We chose to perform the inversions in a column-mass scale. In comparison to optical-depth, column-mass allows computing the gas pressure scale directly, without involving the equation of state or background opacities. We will see later that this feature is important when using the neural network to predict models from observations.

\begin{figure}
    \centering
    \includegraphics[width=\columnwidth,trim=0.5cm 0.2cm 0.5cm 0.8cm, clip]{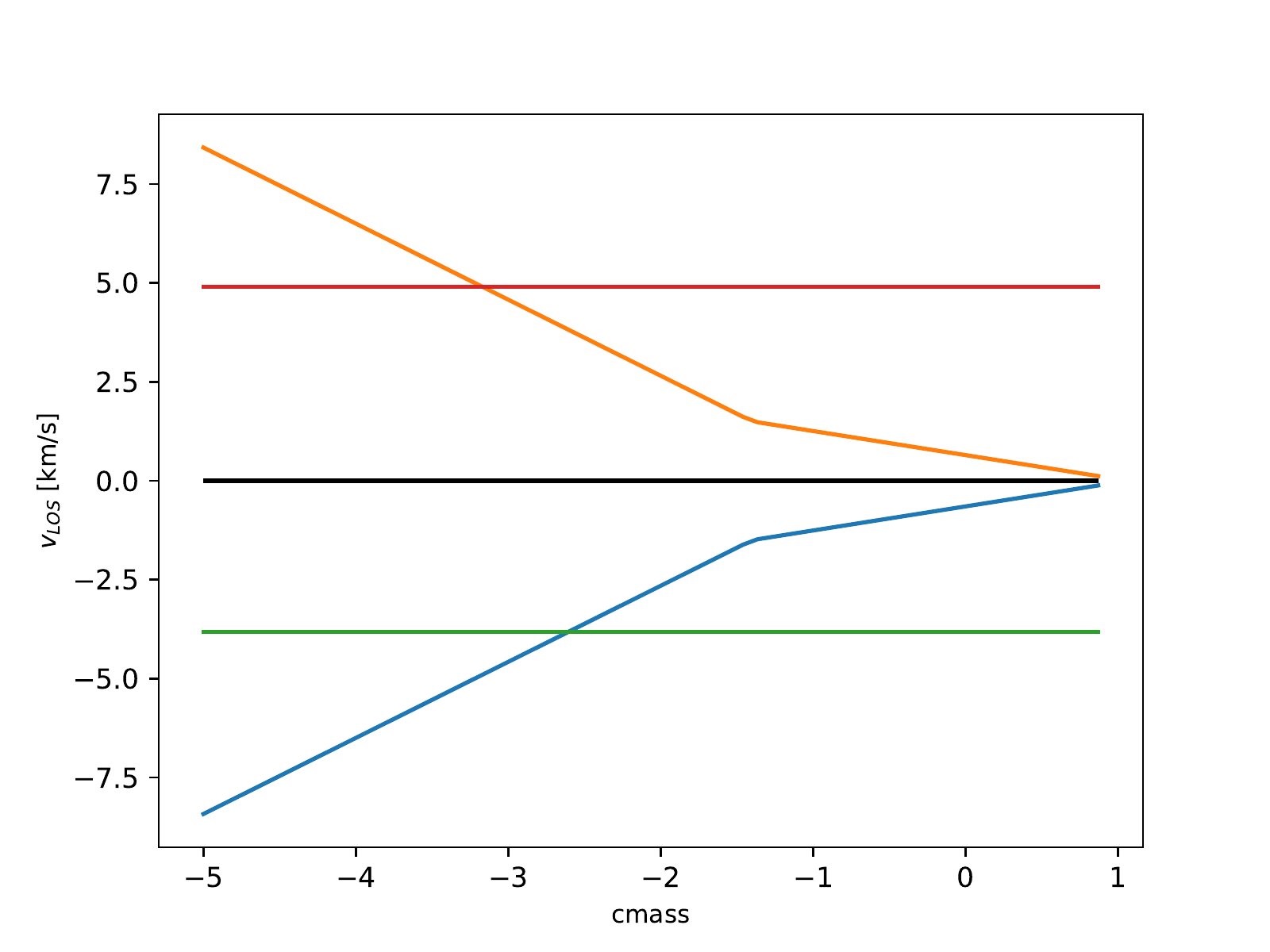} %trim=2.5cm 1cm 2.5cm 0.5cm, clip
        \caption{Example of the line-of-sight velocity stratifications as a function of the logarithm of the column mass ($\xi$), used to create the models for the inversions with STiC.}
    \label{vlos_trend}         
\end{figure}

\begin{figure*}[!ht]
        \centering
        \includegraphics[width=\textwidth]{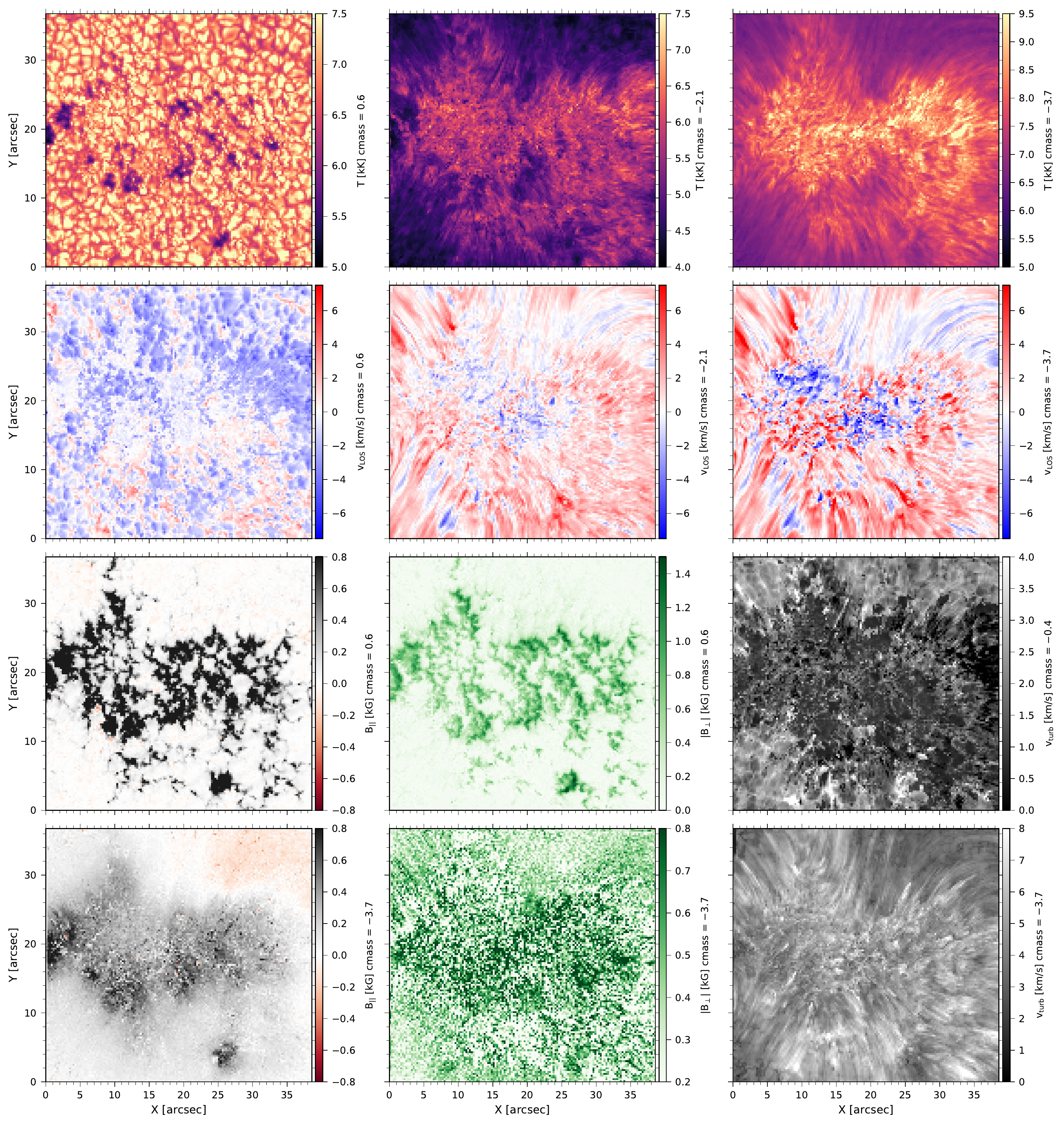}
        \caption{Horizontal cuts from the final model atmosphere obtained from the inversions. The depth in the atmosphere is given in column mass. The $T$ and the $v_{LOS}$ are shown for three different depth, corresponding to lower photosphere, upper photosphere and chromosphere. $B_{||}$, $B_{\perp}$ and $v_{turb}$ are shown for two depth corresponding to lower photosphere and chromosphere. 
         }
        \label{stic_inv}
    \end{figure*}

%\jcr{[Shouldn't we mention that you are selecting a training dataset and therefore it is better to have many pixels spread across the FOV rather than many pixels in a small section of the FOV?]}
We inverted a training set for a neural network. The training set consisted of the full field-of-view for two non-consecutive time steps of the series, but in order to speed up the calculations we only inverted every seventh pixel of the FOV. Furthermore, in order to efficiently train the neural network it is more convenient to have a statistical picture of FOV, with many pixels spread across the region and many different observed spectra, rather than many pixels in a small area that could all present similar observed spectra.

%Instead of starting the inversion from a standard model such as FAL-C model \citep{1993ApJ...406..319F}, and apply random perturbations to the input parameters, we decided to use a different approach. 

We initialized the magnetic field vector of the initial input model using the spatially-regularized weak-field approximation proposed in \citet{2020A&A...642A.210M}.
The presence of strong velocity gradients as a function of depth can greatly distort the line profile of very strong lines. In out tests, this was one of the main sources of degeneracy in the output models. In order to better sample the parameter space in line-of-sight velocity, we prescribed five initializations of the stratification that were added to the FAL-C model \citep{1993ApJ...406..319F}: three with constant values at $0,\pm 5$~km~s$^{-1}$, and two with strong upflowing and downflowing gradients (see Fig.~\ref{vlos_trend}). %Two of the models present a gradient in $v_{LOS}$ that become steeper in the chromosphere, while the other two have a constant velocity for the whole solar atmosphere. Moreover, two of the models are positive and two are negative in order to account for all the possible values. 
The number of nodes used to run the inversion are the same for all five models. We used 10 nodes in temperature, 4 in $v_{LOS}$, 4 in $v_{turb}$, 3 in $B_{||}$, 2 in $B_{\perp}$ and 1 node in $\phi$. 
%Then, we took an average of the results of the five models and merge it in a final model. After increasing the number of nodes in temperature to 11 and the number of nodes in $B_{||}$ to 4, we ran another cycle. 
For each pixel, we selected the model that yielded the best $\chi^2$ value and then applied a mild horizontal smoothing. The inversions were re-started with an increased number of 11 nodes in temperature and 4 nodes in $B_\parallel$. Once all cycles were finished, we switched on the NLTE equation of state and re-run the final cycle, that essentially accounts for the ionization of hydrogen in NLTE and the ionization of the rest of elements is calculated in LTE. The last cycle essentially affects the temperature stratification and the derived electron densities.

%We have then repeated the last cycle of inversions keeping the same number of nodes but changing the settings about non-LTE conditions. Indeed, so far we have run STiC assuming an LTE equation of state, so the ionization balance is done in LTE. Now we changed the inversion set up in order to take into account non-LTE electron densities.

%The obtained final model atmosphere is shown in Fig.~\ref{stic_inv}. In order to speed up the inversion process, we reduced the dimensions of the FOV by skipping 7 pixels for each analyzed pixel, assuming that abrupt changes in the spectra do not occur in neighbour pixels. 
The first two rows in Fig.~\ref{stic_inv} present respectively the $T$ and $v_{LOS}$ for three different depths in the atmosphere, that correspond to lower photosphere, upper photosphere and upper chromosphere. For $B_{||}$, $B_{\perp}$ and $v_{turb}$ just two depths have been chosen: they are represented in each column of the two bottom rows. %The depth in the atmosphere is expressed in column mass since the inversions have been completed using this depth type. 

%The results of the inversions are in line with the observed active region. 
The inversion results reproduce many features of the solar atmosphere that are well known from past research.
In the photospheric temperature map, for example, the granulation pattern is visible, while moving upwards in the atmosphere the hotter and elongated fibrilar structures extend toward the outside. Moreover, the magnetic field presents the typical plage structure with strong concentration of field and field-free gaps in the photosphere, while in the chromosphere it is more extended and less strong \citep{2015A&A...576A..27B,2020A&A...642A.210M}. We note that the $Q$ and $U$ signals in the chromosphere are very noisy and the inversion code struggles to reconstruct clean maps for the $|B_\perp|$ and the azimuthal components of the magnetic field.

Due to the S/N ratio of the observation and to the different sensitivity of the emerging intensities to the different parameters of the model, the depth resolution of the model is greatest in temperature, and line-of-sight velocity, and much more limited in microturbulence and $B_\parallel$. Therefore the magnetic canopy does not appear as sharply in the reconstructed magnetic field stratification as in the temperature reconstruction and in both cases it is much smoother than it probably is in reality because of the limitations of a nodes-based inversion.

\subsection{Neural network application} \label{nn}

%(pytorch_total_params_grad: 316677)
Since it would be extremely time consuming and very computationally expensive to invert all the 35 time frames of the observation with STiC, we suggest instead a easier and faster approach using neural networks. They have showed a good performance in terms of accuracy and speed for solving different problems, for example to identify and predict solar flares \citep{2018ApJ...861...62P}, to denoise solar observational images \citep{2019A&A...629A..99D}, or to study the mapping between spectral lines and the solar atmosphere \citep{2005ApJ...621..545S,2021arXiv211203802C}. In this case, we have used the resulting model atmospheres from the inversion to train a neural network that can quickly predict the model atmosphere for the rest of the dataset \citep{2019A&A...626A.102A}. We refer the reader to Appendix~\ref{ap:nn} for a detailed explanation of the architecture, training process and validation of the neural network.

% . It is one of the most used type of artificial neural network (ANNs) because it is relatively easy to structure but at the same time it is well performing. For more details, \citet{Schmidhuber_2015} presents an overview of the main characteristics of this type of neural network.

\begin{figure*}[!ht]
        \centering
        \includegraphics[width=\textwidth]{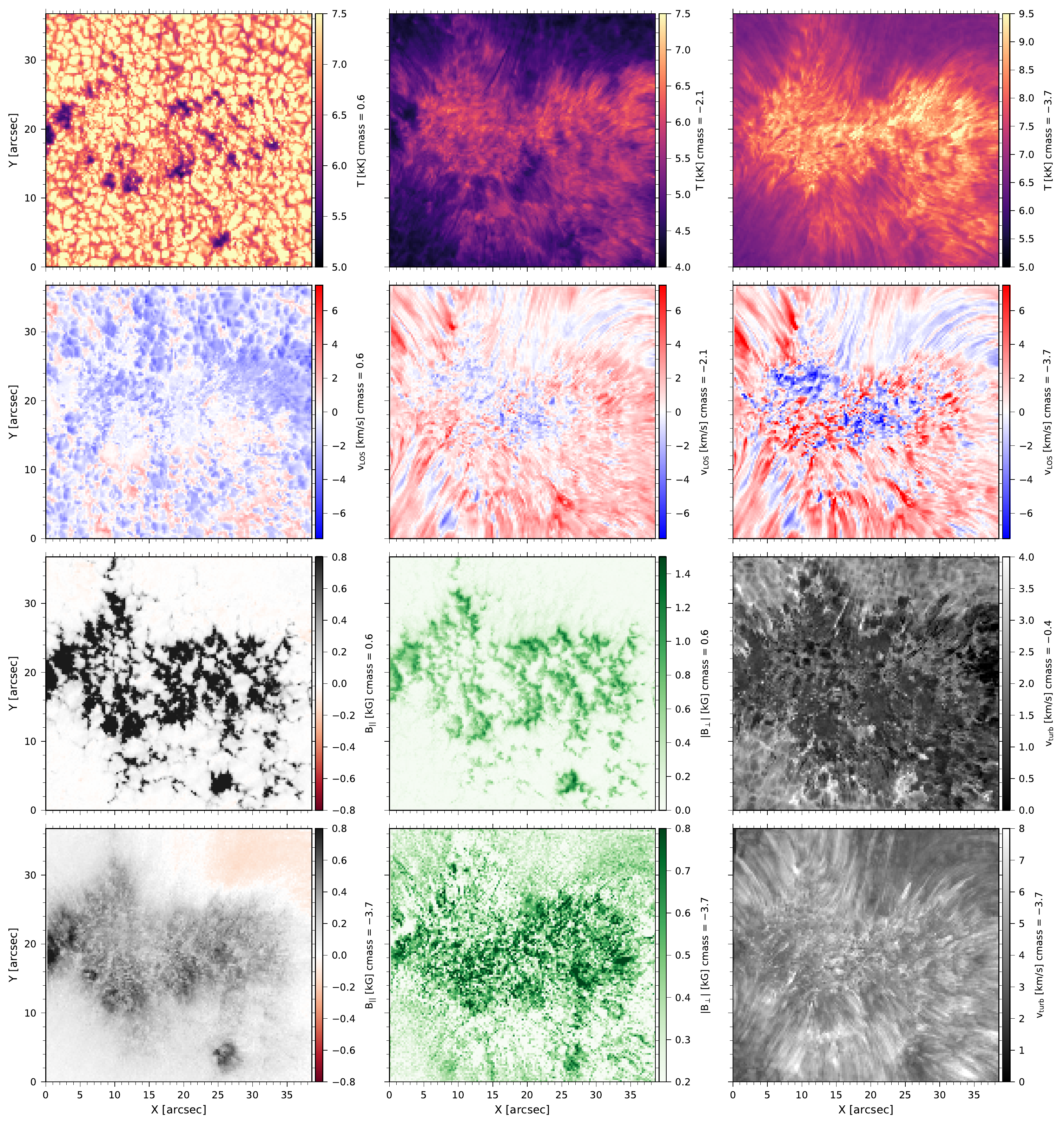}
        \caption{Final model atmosphere of the field of view obtained from the training of the NN. Format as for Fig.~\ref{stic_inv}.
         }
        \label{nn_inv}
    \end{figure*}
\begin{figure*}[!ht]
        \centering
        \includegraphics[width=\textwidth,trim=0cm 0.1cm 0.1cm 0cm, clip]{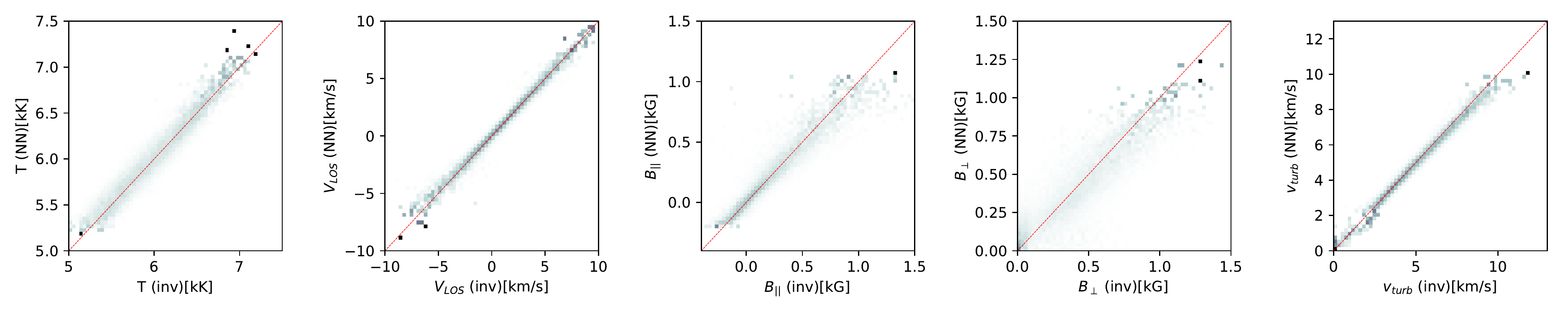}
        \caption{Density plots comparing the results of the inversion with the results of the neural network. \emph{From the left:} The panels show the correlations for $T$, $v_{LOS}$, $B_{||}$, $B_{\perp}$ and $v_{turb}$ at $cmass=-3.1$.
         }
        \label{scatter}
    \end{figure*}  

\paragraph{Results of the NN:} The results obtained with the training of the NN are shown in Fig.~\ref{nn_inv} in a format similar to Fig.~\ref{stic_inv}. The represented time step is the same as in Fig.~\ref{stic_inv}.

The temperature and line-of-sight velocity prediction from the NN are very well correlated with the results from the inversions (see 2D density plots in Fig.~\ref{scatter}). Both components of the magnetic field vector are predicted smoother compared to those obtained with the inversions. The latter occurs because the noise from Stokes $Q$ and $U$ was dominating the inversion results and STiC was not able to fully reconstruct the signal. The introduction of a constant factor in the training helps the neural network to smooth the noise and to better estimate the values.

The NN seems to overestimate $v_{turb}$ very deep in the photosphere, while in the chromosphere this behaviour disappears. The neural network incorrectly correlates areas of strong photospheric $B_{||}$ with higher values of the microturbulence. We investigated ways of removing this degeneracy, but we could not find a solution. We decided to move forward regardless because the values of the microturbulence close to the continuum formation layer in the photosphere have no influence in the prediction of radiative losses through strong chromospheric lines. %is layer of the solar atmosphere is not included in our calculations and the correlated errors could be neglected.
%\jcr{[Does the Vturb prediction look better around cmass = -0.3? The panel shown in Fig. 3 is extracted at the lower boundary of the atmosphere, I don't think we have sensitivity to vturb in that layer that is deeper than the continuum formation height. What you see could be an extrapolation of the results a bit higuer up.}

\section{Radiative cooling rates}\label{sec:radiative}
%The energy balance of a specific layer of the solar atmosphere can be estimated through the calculation of the radiative cooling rates. They represent the fraction of the energy that needs to be restored in order to sustain the atmosphere. 
The heating terms in the chromosphere can be approximately estimated by calculating the integrated radiative losses, since the energy required to sustain radiative losses must be sustained by chromospheric heating terms.
To calculate the radiative cooling rates from the predicted model atmosphere from the NN, we first imposed hydrostatic equilibrium in order to derive a z-scale and the gas pressure scale.
%From that, we were able to calculate the gas pressure from the relation between the latter and the solar gravity. 
Most inversion codes operate in an optical-depth scale and therefore the gas pressure has to be calculated iteratively with the consequent re-calculation of the continuum opacity \citep[see, e.g.,][]{1970stat.book.....M}. However, working in a column mass scale, simplifies the calculations and no iterations are needed \citep{2014tsa..book.....H}:
\begin{equation}
    p_{gas}= g_{\odot}\xi
\end{equation}
where $\xi$ is the column mass, known from the inversions, $g_{\odot}$ is the solar gravity and $p_{gas}$ is the gas pressure.

We have calculated the electron densities in each depth point using a simple equation of state\footnote{https://github.com/jaimedelacruz/pTau} proposed in \citet{1970stat.book.....M} which includes hydrogen atoms bound in $H^-$ and $H2$ molecules. Once the electron densities were known, we estimated the total number of atoms for a given temperature assuming an ideal gas: 
\begin{equation}
    p_{gas} = (N_{a} + N_{e})  K_{B}  T,
\end{equation}
where $N_{a}$ is the atoms number density, $N_{e}$ is the electron density, $K_B$ is the Boltzmann constant and $T$ the temperature. In this way, it is possible to obtain the total number of atoms $N_{a}$ and, by multiplying it to the mean particle mass, we can obtain the density $\rho$ at a certain depth in the atmosphere: 
\begin{equation}
    \rho = <m>N_a,
\end{equation}
where the mean particle mass $<m>$ is given by:
\begin{equation}
    <m> = \sum_i^{N_{elem}} (a_i  m_i).
\end{equation}
In this case, $a_i$ is the solar abundance of the $i$-th element and $m_i$ is its atomic mass. 

Finally, we obtained the z-scale from the definition of column mass:
\begin{equation}
    \xi= \int \rho dz
\end{equation}
therefore, in a discrete grid we can write:
\begin{equation}
    dz_{i} = \frac{\xi_i-\xi_{i-1}}{(\rho_i+\rho_{i-1})/2}.
\end{equation}
This \emph{hydrostatic} z-scale is used to calculate the integrated radiative losses in the chromosphere. The latter are calculated integrating over a height interval, selected from the z-scale. Our final model contains all the quantities calculated under the assumption of hydrostatic equilibrium and the ones previously obtained after the inversion process.

%In synthesis mode, STiC is able to read an input model and to provide as output files the synthetic spectra and an \emph{extra} file with all the information about the atom that we need for our purposes. It computes the upper and lower levels at which the transitions happen, the collisional and radiative rates, the population of each level, the radiative losses and the electron densities. 
We have used a modified version of STiC that evaluates directly Eq.~(\ref{loss_integral}) and also outputs the net radiative rates for all transitions and the atom population densities. In our case we considered Ly$\alpha$, Ly continuum, H$\alpha$, the \ion{Ca}{ii}~H\&K and the IR triplet lines and the \ion{Mg}{ii}~h\&k and UV triplet lines. Our inversion setup includes lines that sample the solar atmosphere from the upper chromosphere to the photosphere. Although the \ion{Ca}{ii}~K line may be sensitive to the very lower part of the transition region, it is not sufficient to properly constrain its exact location and gradient. Therefore, the reconstructed transition region is very cold and extended in comparison to models from inversions including transition region diagnostics \citep[see e.g.,][]{2016ApJ...830L..30D}. The latter seem to have a large impact in the prediction of the Balmer continuum, which is unrealistically large in our calculations. We could not find an obvious solution to this problem, and therefore we did not include the Balmer continuum contribution in our study. The effect of this exclusion is that our radiative losses will potentially be even lower than the heating terms that we are approximating with them.
%Indeed, the transitions that contribute more in the energy balance in the chromosphere are those from \ion{Ca}{ii} infrared triplet, \ion{Ca}{ii} {H\&K}, \ion{Mg}{ii} {h\&k} and \ion{H}{} \citep{1981ApJS...45..635V}. 

The inversions were calculated including the effect of NLTE hydrogen ionization in statistical equilibrium by imposing charge conservation \citep{2007A&A...473..625L}. But our NN does not predict the electron density. Therefore, we have first calculated a forward synthesis with the \ion{H}\ atom, recovering not only the radiative losses but also the electron densities in NLTE. Then the electron densities in the original model have been replaced with the new ones in the input model. The synthesis was done for the \ion{Ca}{ii} and \ion{Mg}{ii} atoms with the updated electron densities. 
%In order to proceed with the synthesis of an atom rather than another, the input atoms file, that contains information about active and passive atoms, needs to be modified in each cycle. In this way, only one atom in each cycle is active, while the other passive atoms only contribute to the background opacity.

The radiative cooling rates are calculated automatically inside the code by computing the divergence of the radiative flux \citep[see e.g.,][]{2002ApJ...565.1312U,2003rtsa.book.....R}:
\begin{equation}
Q=  \nabla \cdot F = \int_{0}^{\infty} \alpha_{\nu}(z) \left [ S_{\nu}(z) - J_{\nu}(z)\right ] \, \mathrm{d}\nu,
\label{loss_integral}
\end{equation}
where $\alpha_\nu$ is the total absorption coefficient, $S_\nu$ is the total source function and $J_\nu$ is the mean intensity over solid angle. For bound-bound transitions, Eq.~(\ref{loss_integral}) can be transformed into an expression that only depends on the net radiative rates and the level population densities:
\begin{equation}
    Q =h \nu_0 (n_u R_{ul} -n_l R_{lu}),
\end{equation}
where $h$ is the Planck's constant, $\nu_0$ is the central frequency of the transition, $n_{u/l}$ are the population of the upper and lower level respectively and $R_{ul/lu}$ are the radiative rate coefficient from the upper to the lower level or vice versa. 

In order to obtain the integrated radiative losses, $Q$ has to be integrated over the range of geometrical heights of the region of interest. In our case the integration limits are set in order to take into account only the chromosphere, so from the depth point after the temperature minimum to the depth point at which the temperature reaches $T\sim 10000$K. We note that moving the integration limits can change the derived losses, and therefore some deviations can be expected when comparing numbers from different studies. 

%\jcr{[Why not integrate from 10000~k to the temperature minimum (or a few grid points above the temperature minimum) in every pixel?]}

\section{Results}
\subsection{Single timestep}
\begin{figure}[!ht]
        \centering
        \includegraphics[width=\columnwidth]{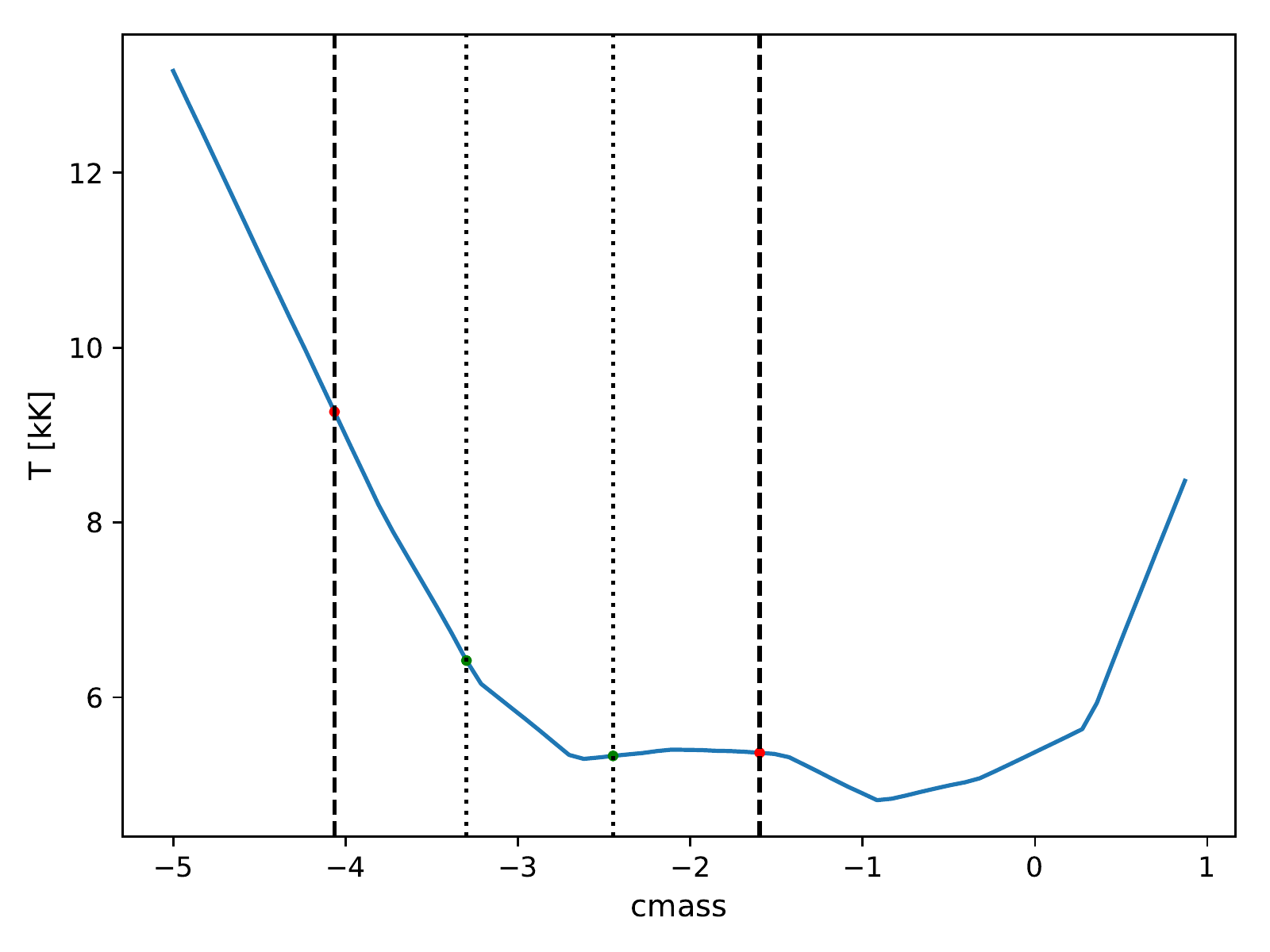}
        \caption{Average temperature for the plage region as a function of $cmass$. The red dots and the dashed lines represent the total integration interval for the radiative losses, while the green dots and the dotted lines show the division between \emph{lower}, \emph{middle} and \emph{upper} integration interval used in Fig.~\ref{Q_contour} and Fig.~\ref{Q_layer}.
         }
        \label{int_interval}
\end{figure}

\begin{figure*}[!ht]
        \centering
        \includegraphics[width=\textwidth]{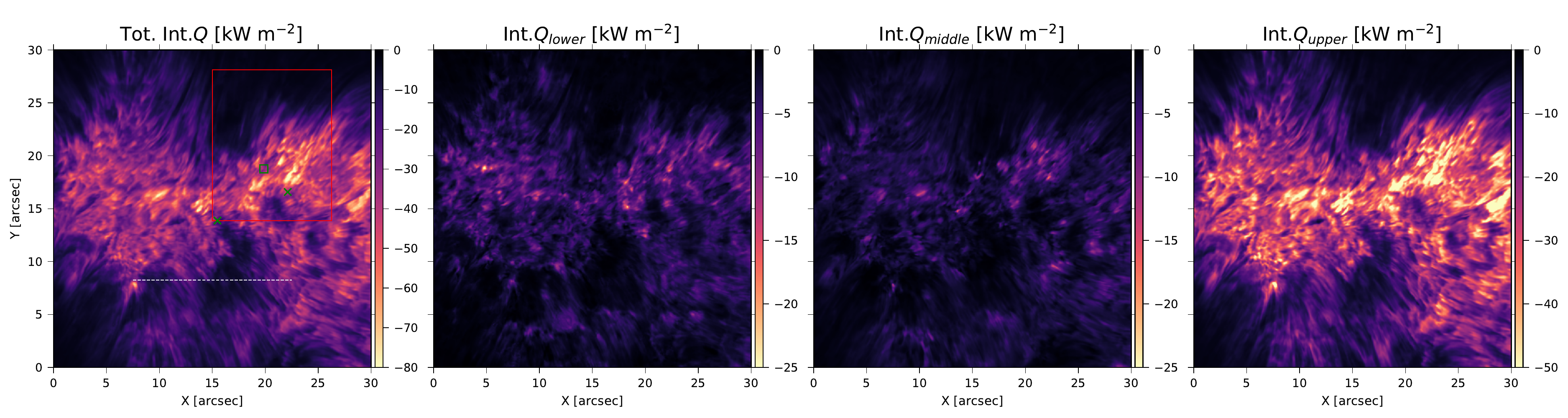}
        \includegraphics[width=\textwidth]{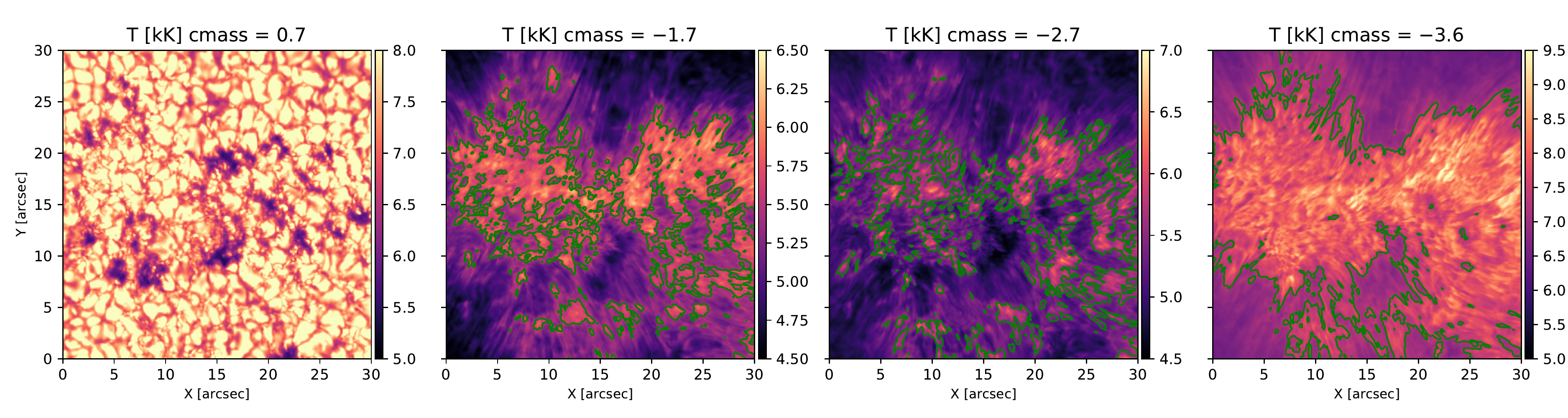}
        \includegraphics[width=\textwidth,trim=0cm 0.1cm 0.39cm 0.22cm,clip]{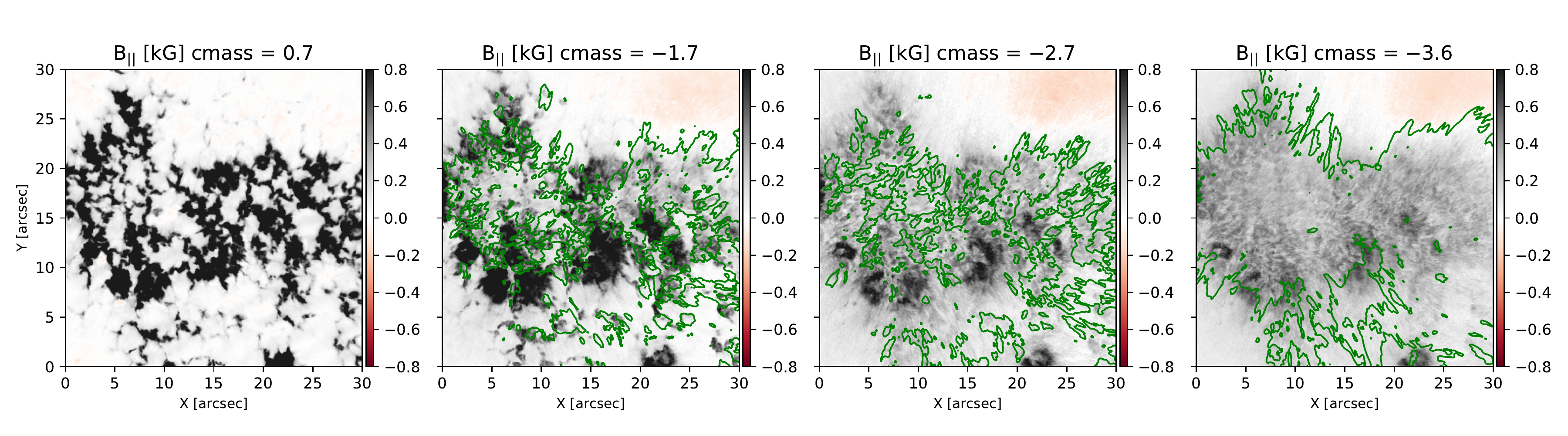}
        \caption{\emph{Top row:} derived radiative losses for the entire FOV for the first time step. The first panel from the left represents the total radiative losses integrated over the chromosphere. The other three panels show, respectively, the radiative losses integrated over the lower, middle and upper chromosphere. \emph{Middle row:} Maps of the temperature for four different heights in the solar atmosphere. The height is increasing for left to right. \emph{Bottom row:} Maps of the parallel magnetic field for four different heights in the solar atmosphere. The height is increasing from left to right. The first panels in the middle and bottom rows represent respectively the $T$ and $B_{||}$ in the photosphere. The green contours indicate the area where $Q< const*\tilde{Q}_{layer}$ kW m$^{-2}$ in the corresponding atmosphere layer, where $\tilde{Q}_{layer}$ is the median value of $Q$ in the corresponding layer. $Const=1.6$ for the lower and middle chromosphere, $const=0.7$ for the upper layer.
         }
        \label{Q_contour}
    \end{figure*}

In order to study the spatial distribution of radiative losses, we have calculated the radiative losses for the initial time step of the observations. We have chosen this frame because it is one of those with better seeing conditions. We have divided our integration interval (in height) into three subregions (lower, middle, upper chromosphere) to understand how the energy deposition is taking place, which are shown in Fig.~\ref{int_interval}. The lower extreme of the interval has been chosen after the temperature minimum, that define the end of the photosphere and the beginning of the chromosphere. The upper extreme has been chosen in order not to include the transition region in the calculations. Since our inversions did not include lines that are strongly sensitive to the transition region, the steep temperature gradient and its exact location are not well constrained in our models.%the estimation of the radiative losses for this area is not in the purpose of this study.

\begin{figure*}[!ht]
        \centering
        \includegraphics[width=\textwidth]{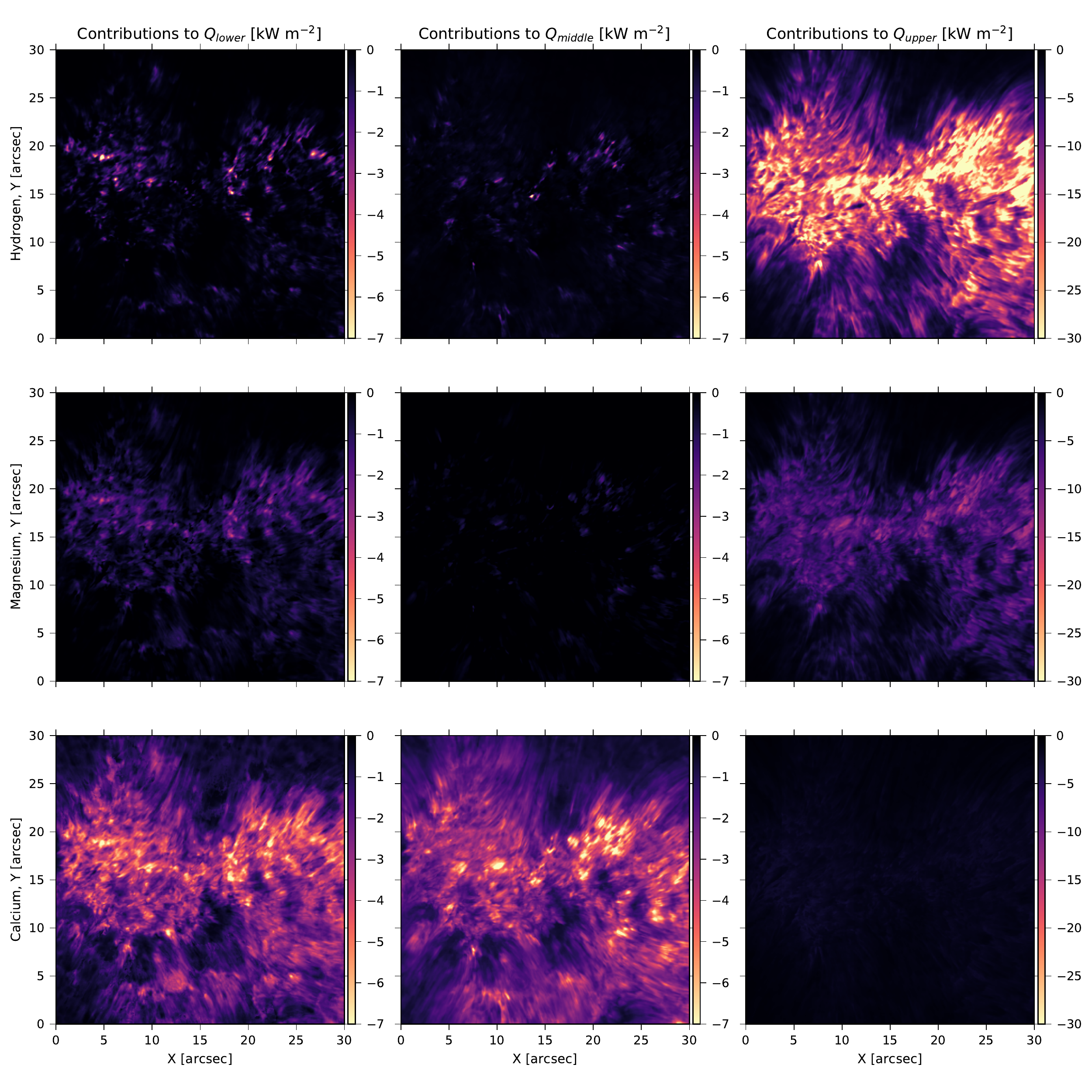}
        \caption{Maps of the contributions to the obtained radiative losses divided by lower, middle, upper layer in the chromosphere and divided by atom. \emph{Top row:} Integrated radiative losses for Hydrogen. \emph{Middle row:} Integrated radiative losses for Magnesium. \emph{Bottom row:} Integrated radiative losses for Calcium.
         }
        \label{Q_layer}
    \end{figure*}

The derived radiative losses integrated over the entire interval and over the three subregions of the chromosphere, are shown in the top row of Fig.~\ref{Q_contour}. The average integrated radiative losses over the whole plage region are $\sim -28$~kW m$^{-2}$. The FOV shows very small-scale regions with peak values of $\sim -90$~kW m$^{-2}$. 
In order to have a more precise insight about the calculated radiative losses, we have plotted in Fig.~\ref{Q_layer} the contributions of each atom in the tree sub layers of the chromosphere. The \ion{Ca}{ii} contribution is dominating in the lower and middle chromosphere while the contributions from \ion{H}{i} and \ion{Mg}{ii} lines are negligible in this region. The middle chromosphere is the sub layer that presents the lower values associated with the radiative losses.
%Jcr: We just said the same before!: The dominant transitions are the \ion{Ca}{ii} infrared triplet and \ion{Ca}{ii} {H\&K}. 
%As the height increases the contribution from the Hydrogen starts to be more consistent. 
%\jcr{The \ion{H}{i} dominates in the upper chromosphere.}
%, the largest amplitudes are located in very small-scale features and they are part of the \ion{H}{i} contribution. These features are co-spatial with the strongest signals in the lower chomosphere}. 
%\jcr{[Jaime: are you including the Balmer continuum? You mention "transitions (plural) from the Balmer". Otherwise mention simply Halpha.]}
%The bound-bound transitions from the Balmer series are the ones that contribute most. 
In the upper chromosphere, Hydrogen is the main contributor to the radiative losses. The contribution from the  \ion{Mg}{ii} atom is approximately one half of that from Hydrogen. %In this layer, also $Ly\alpha$ is also one the major transitions, together with \ion{Mg}{ii} {h\&k}. These results agree well with previous studies \citep{1985cdm..proc...67A}.

The comparison of the integrated radiative losses with the line-of-sight component of the magnetic field provides great insight into the overall heating process (see Fig.~\ref{Q_contour}). From left to right, the contours plotted in three of the four panels correspond to a fraction of the median radiative losses value ($Q<const*\tilde{Q}_{layer}$ ~kW~m$^{-2}$, where $\tilde{Q}_{layer}$ is the average net radiative loss in that layer) in the lower chromosphere, in the middle chromosphere and in the upper chromosphere. In the lower and middle chromosphere the bulk of the radiative losses are concentrated in the areas surrounding the strongest photospheric magnetic field concentrations but not inside the latter. 
%In those locations is precisely where \citet{2013ApJ...764L..11D} found \emph{magnetic canopy} profiles in the 8542~\AA\ line and where \citet{2020A&A...642A.210M} reconstructed the magnetic canopy effect. 
The distribution of the largest temperatures is also greatly correlated with the magnetic canopy. The photospheric temperature panel shows that our FOV contains a number of small pores. We will discuss the effect of pores in \S\ref{sec:pores}. %The same panel also shows contours with the peak values of the radiative losses in the lower chromosphere, and we see that they are located very close to the photospheric footpoints of the magnetic field. 
In the lower chromosphere, the peak values of the radiative losses reach $\sim -20$~kW m$^{-2}$. The lower-left panel in Fig.~\ref{Q_layer} shows that the \ion{Ca}{ii} contribution dominates in the lower chromosphere, and the canopy shape is already visible there.

In the middle chromosphere the magnetic field becomes smoother and the magnetic canopy is clearly visible in the $B_\parallel$ image, suggesting that at this depth we are already sampling above the lower edge of the canopy. The $Q_{middle}$ shows a similar picture with slightly smaller radiative losses. %These finely structured features are surprising when compared to the relatively smooth and homogeneous magnetic field map, but we will return to these features in the analysis of the time-series (\S\ref{sec:time}). 
The temperature image shows a nearly homogeneous value of approximately 6.5~kK, while most pores appear as cold holes in the canopy.

In the upper chromosphere, the integrated radiative losses are dominated by the \ion{H}{i} contribution. In this layer the radiative losses, the enhanced chromospheric temperature and the magnetic field form a patch above the plage target with relatively constant values of $<Q>\approx -22$ kW~m$^{-2}$, $<T> \approx 8.5$~kK and $|B_\parallel| \approx 370$~G. In this layer only the strongest pores are visible in the temperature map and in the radiative losses map.

\begin{figure*}[ht!]
    \centering
    \includegraphics[width=5.22cm, trim=0.2cm 0.1cm 0.1cm 0.2cm, clip]{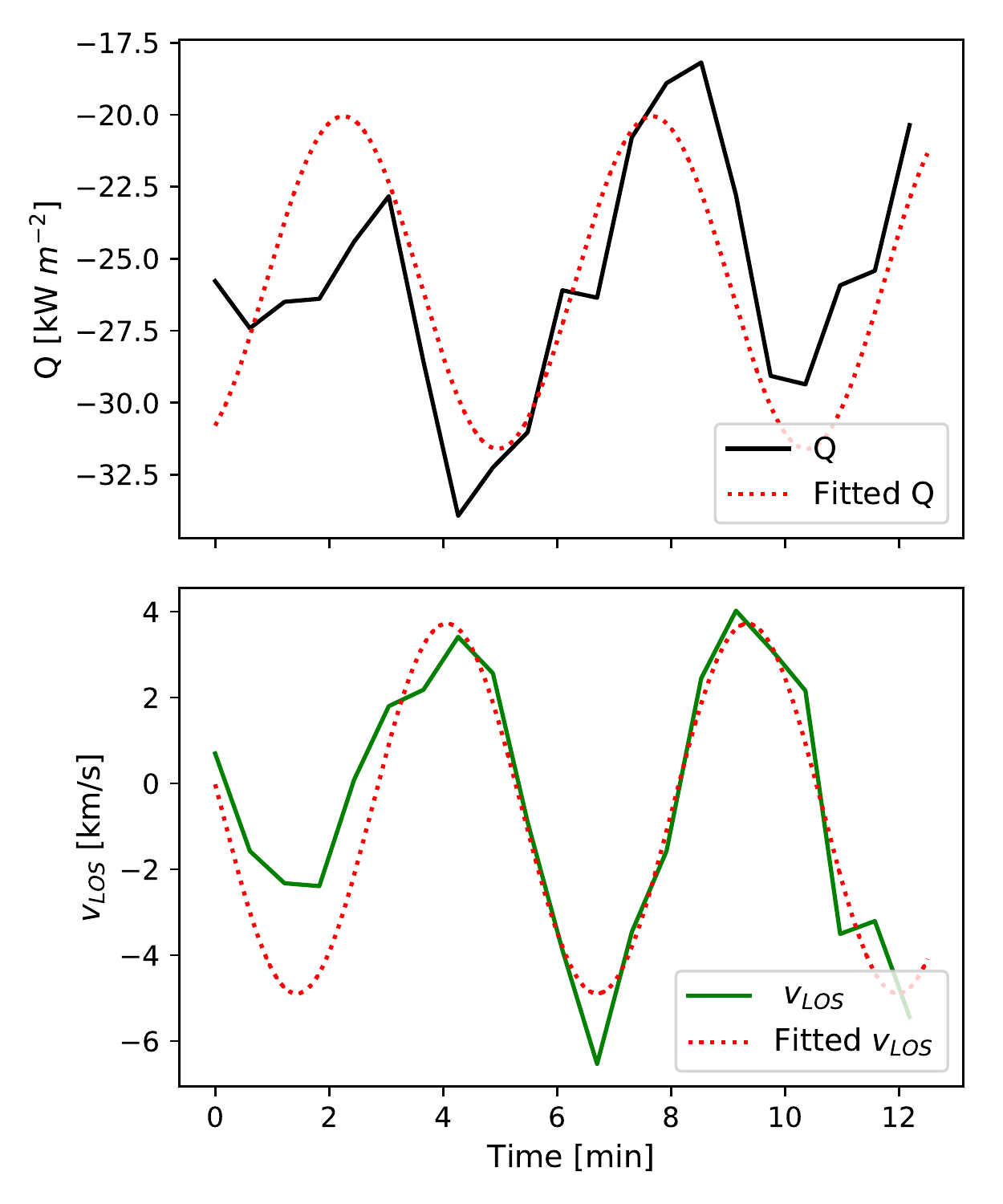} \hfill
    \includegraphics[width=5.22cm, trim=0.15cm 0.1cm 0.1cm 0.2cm, clip]{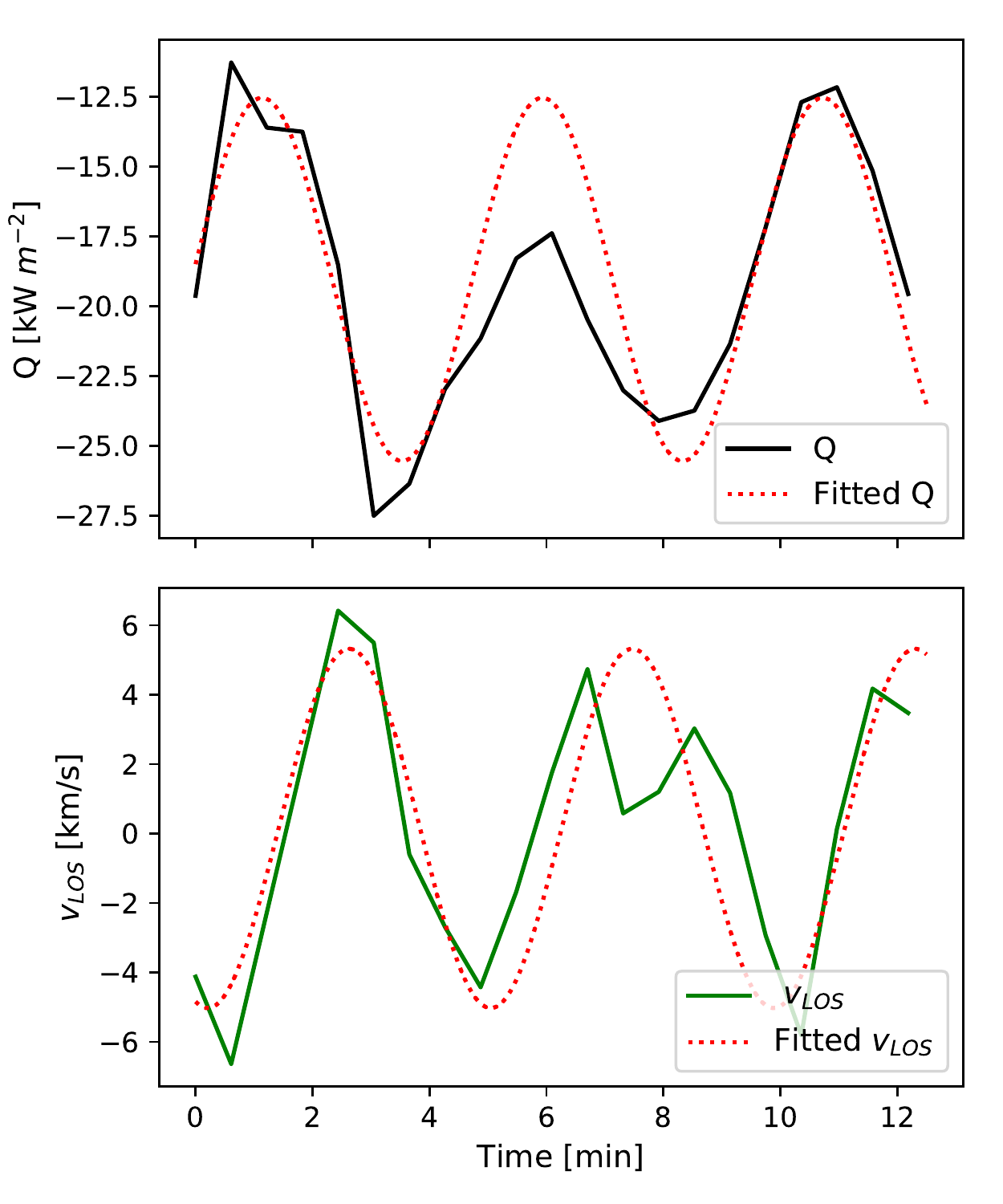} \hfill
    \includegraphics[width=7.79cm, trim=0.7cm 0.cm 0.5cm 0.5cm, clip]{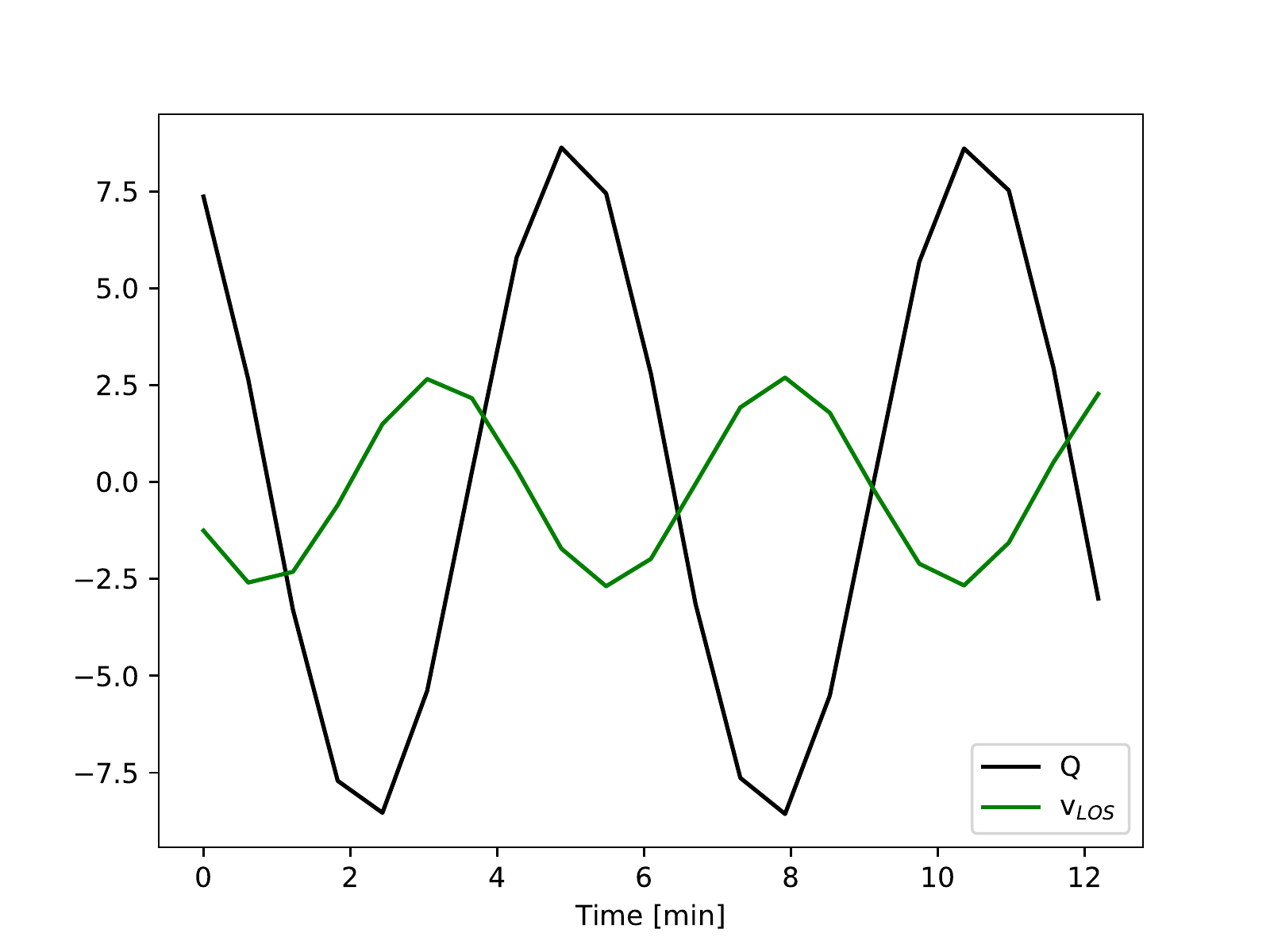}
    \caption{Time evolution of the radiative losses and of $v_{LOS}$ for two selected pixels and a region in the FOV. The two chosen pixels are indicated with green crosses in Figure~\ref{Q_contour}, while the region is indicated with green boundaries. \emph{Left:} Time evolution of $Q$ (\emph{top panel}) and $v_{LOS}$ (\emph{bottom panel}) for the first random pixel. \emph{Centre:} Time evolution of $Q$ (\emph{top panel}) and $v_{LOS}$ (\emph{bottom panel}) for the second random pixel. \emph{Right:} Average functions over the green region, one for $Q$ (blue) and one for $v_{LOS}$ (green). The offset of each function has been excluded for an easier comparison of the results.}%
    \label{sin_pixels}%
\end{figure*}

\subsection{The effect of pores}\label{sec:pores}
\begin{figure}[ht!]
        \centering
        \includegraphics[width=\columnwidth,trim=0cm 0cm 0cm 0cm, clip]{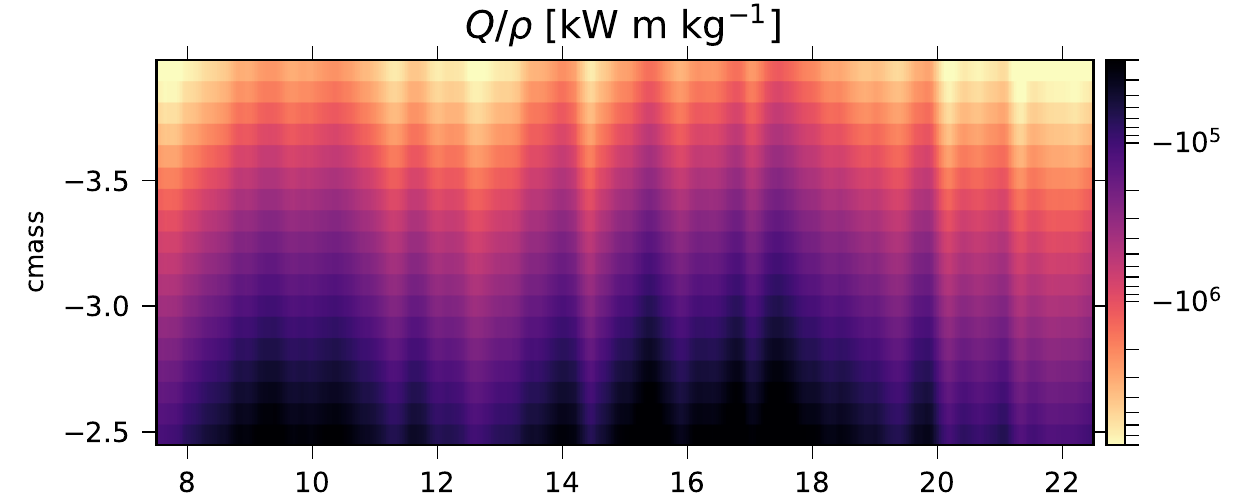}
        \includegraphics[width=\columnwidth,trim=0cm 14.6cm 0cm 14.3cm, clip]{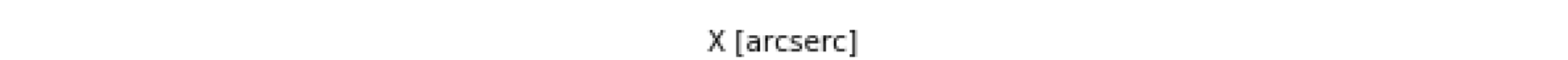}
        \caption{Vertical cut corresponding to the white dashed line of Figure~\ref{Q_contour}. The total integrated radiative losses have been divided by the model densities $\rho$ at and represented using a \emph{symmetrical logarithmic} scale.
         }
        \label{Q_slice}
    \end{figure}

The presence of pores in plage seem to have a clear imprint in the statistics of the derived physical parameters \cite[see e.g.,][]{2021ApJ...906...83C}. Our target contains several pores, that appear as colder and larger footpoints of the magnetic canopy than in the rest of the magnetic elements. But otherwise, the magnetic field is still strongly vertical. The imprint of pores is clear. Because they are much colder than the smaller \emph{bright} flux tubes, the radiative losses are insignificant in comparison with the surroundings until we reach the upper chromosphere. Figure~\ref{Q_slice} shows a vertical reconstruction, marked with the white slit in Fig.~\ref{Q_contour}. This slice cuts through two pores embedded in the plage region (at $Y \sim 8$ arcsec and $8<X<22$ arcsec), illustrating this effect. The figure also suggests that eventually, the imprint of pores is much smoother and weaker in the upper chromosphere.

Summarizing, having pores in the the FOV does affect the derived radiative losses, especially in the lower chromosphere, because the atmosphere is much colder than in regular flux tubes. Besides that, having spatially resolved maps greatly helps to separate their influence from the rest of the FOV.

%Using the densities from the model and the obtained integrated radiative losses for the FOV, we have made a vertical cut along the white line of Fig.~\ref{Q_contour}. The results are shown in Fig.~\ref{Q_slice}, where, the ratio between $Q$ and $\rho$ has been represented in a \emph{symmetrical logarithmic} scale in order to allow both positive and negative values. The used densities correspond to the integration interval of the radiative losses, shown in Fig.~\ref{int_interval}. The vertical cut shows a stratification of the energy equivalent to what we have obtained in the first row of Fig.~\ref{Q_contour}. It can be noticed the presence of two holes, at $X \sim 100$ arcsec and $150<X<180$ arcsec, in which the radiative losses are close to zero. These areas are also colder and start to lose energy only in the highest layers of the atmosphere.

\subsection{Time-series analysis}\label{sec:time}

The NN has made it possible to obtain the model atmosphere for all the time steps of the observation, and thereby, it has allowed us to estimate a time series of integrated radiative losses in the chromosphere over the entire FOV. 
%Therefore, it was easier to calculate with STiC the energy balance for more time steps. 
We have focused on the red region highlighted in the top left panel of Fig.~\ref{Q_contour} and we have calculated the radiative energy balance for 21 time steps of the whole observations. Given the cadence of the CRISP instrument, the latter cover a total time of $\Delta t =  12.19$ min. %In this time interval, it would be possible to detect the time evolution of the radiative losses connected to wave phenomena. 
In \S\ref{sec:intro} we mentioned that magnetoacoustic waves and shocks can have a significant contribution to the heating of plage, and their imprint should be periodic.
Our aim is to separate the contribution of Ohmic heating from the contribution of waves/shocks \citep{2007ApJ...655..624D,2008ApJ...680.1542H} by analyzing a time series. 
%Indeed, these are among the physical processes that could explain the predicted radiative cooling rates \citep{2008ApJ...680.1542H, 2011ApJ...736....3V}.

%\sout{The radiative losses for each pixel in the selected FOV have been plotted as a function of time}\jcr{[Where have you plotted them?]}. They are calculated integrating over the whole interval, between $cmass \sim -2.5$ and $cmass \sim -4$. After noticing a sinusoidal behaviour, we fitted the curves with a sinusoidal function. 
In order to have a grasp on the dominant period ($p$), amplitude ($A$), phase ($\phi$) and offset ($C_{\mathrm{off}}$) of the oscillatory behaviour, we have fitted a very simple model $y = C_{\mathrm{off}} + A\cdot\sin(\phi +p \mathrm{t})$ to the $v_{\mathrm{LOS}}$ and $Q_{\mathrm{tot}}$ temporal curves. The results for two random pixels are shown in the left and in the middle panels of Fig.~\ref{sin_pixels}. The selected pixels are marked with green crosses in Fig.~\ref{Q_contour}. The $v_{LOS}$ is estimated in the region at $cmass= -3.8$, corresponding to middle/upper chromosphere. The results are in line with previous studies about waves propagation in different targets of the solar atmosphere.
%We have also compared the behaviour with the evolution of $v_{LOS}$ over time noticing the same sinusoidal trend, but with a phase difference. 
Both curves have very similar periods, but there is a phase shift between $Q_{\mathrm{tot}}$ and $v_{\mathrm{LOS}}$, which we discuss below.

%In a recent study focused on sunspot umbrae, \citet{2021ApJ...918...47F} found an oscillatory behaviour between temperature and velocity calculated from spectropolarimetric simulations of \ion{Ca}{ii}~8542~\AA. Instead, we computed the energy balance that is anyway strongly correlated to the temperature.

In order to reduce the noise inherent to a single pixel measurement, we selected the green region of Fig.~\ref{Q_contour} and computed the average values of the parameters of the sinusoidal function. The area covers 20x20 pixels, considering that an average over a bigger region would lead to a loss of information as the temporal coherency of the curves is not achieved over large spatial scales. The two sinusoidal averaged functions are out of phase, as in the panels for the specific pixels, with a phase difference of $\Delta \phi \sim 138 \degree$. The periods of the sinusoidal functions of $Q_{\mathrm{tot}}$ and $v_{\mathrm{LOS}}$ are respectively $p_{Q}=5.4$ min and $p_{v_{LOS}}= 4.8$ min. The offsets have been removed from the representation of the average functions for an easier comparison between the two results.

\begin{figure*}
    \centering
    \includegraphics[width=\textwidth, trim=0.1cm 0.1cm 0cm 0.1cm, clip]{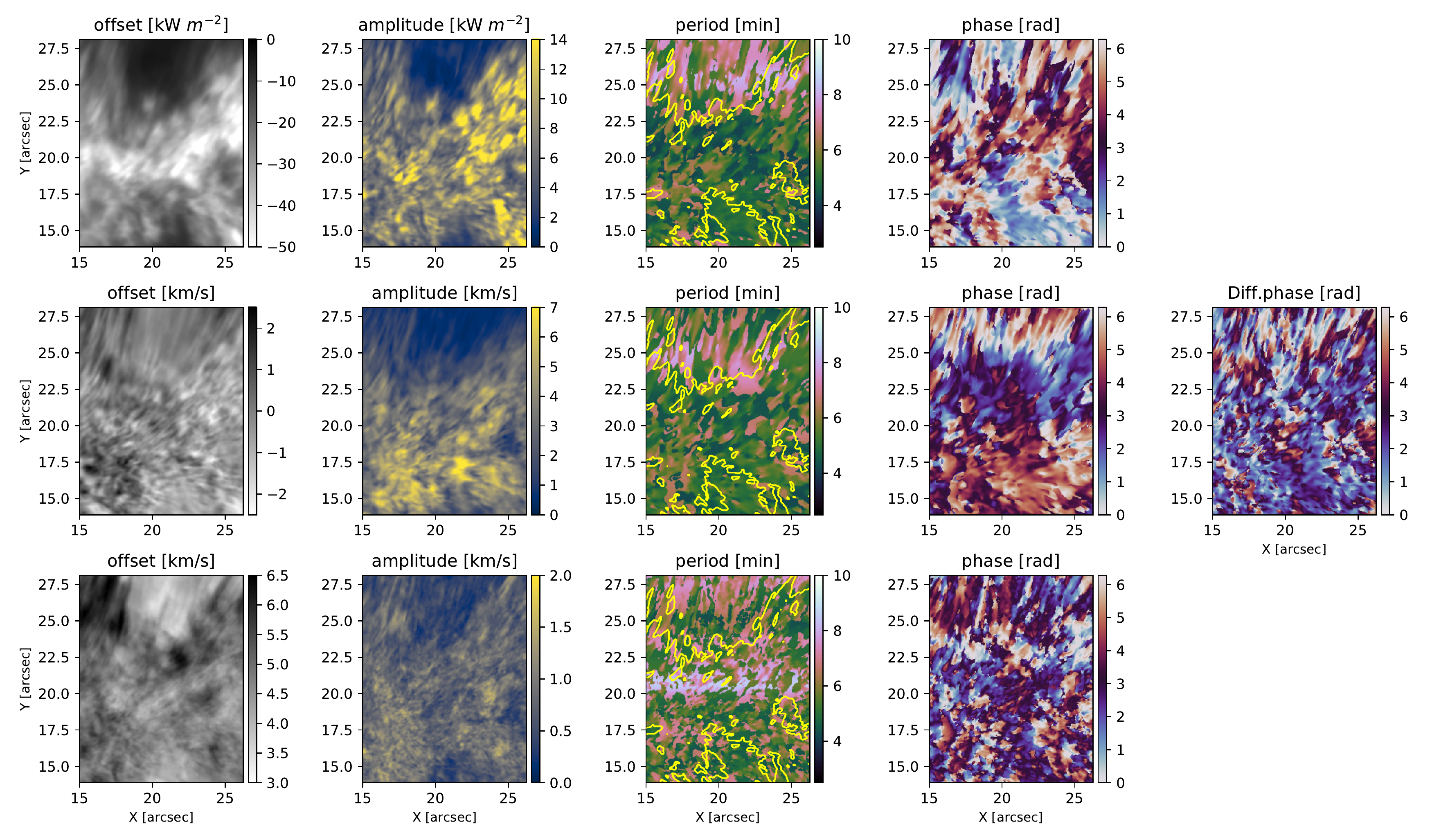}  %trim=2.5cm 1cm 2.5cm 0.5cm, clip 
    \caption{Maps of the parameters (offset, amplitude, period and phase) of the sinusoidal functions obtained for the red region of Figure~\ref{Q_contour}. The top row shows the quantities characterizing the sinusoidal time evolution of the radiative losses, while the second row represents the quantities characterizing the time evolution of $v_{LOS}$ at $cmass=-3.8$. In the maps of the period, yellow contours indicate the plage area corresponding to $Q<-20$ kWm$^{-2}$ (see Fig.~\ref{Q_contour}). The first column from the right shows the phase difference between $Q$ and $v_{LOS}$. The bottom row shows the quantities of the sinusoidal evolution for $v_{turb}$ at $cmass=-3.8$.}
    \label{sin_map}         
\end{figure*}

The fitting procedure has been extended to the red highlighted region in the left top panel of Fig.~\ref{Q_contour}. We have plotted a map for each parameter of the sinusoidal function in Fig.~\ref{sin_map}. The displayed parameters represent the quantities characterizing the time evolution of $Q$, $v_{LOS}$ $and v_{turb}$ (from the top row). As a reference, average values of the parameters over the region are given in Table~\ref{table:1}. 
In our model, the offset is the value of $Q$ or $v_{LOS}$ not related to the periodic wave. In the case of the radiative losses, our interpretation is that it contains the contribution from other heating phenomena, such as Ohmic dissipation, ion-neutral collisions, etc. The top left panel of Fig.~\ref{sin_map} shows a smooth distribution of the offset values. Although the smallest scales are not present in this plot, there is a large scale variation across the panel.% and it is also possible to recognize a structure similar to the one of the radiative losses in the same area. 
%\jcr{The map of offset values turns out to be relatively smooth, and that is also a clue that the main heating source could be related to magnetic-field regulated processes. The magnetic field is expected to be smooth and room-filling in the upper chromosphere, above the $\beta=1$ layer where magnetic forces dominate over those arising from gas pressure gradients.}%The central part of the region, where the plage is extending, is characterized by lower values, as in the map of the radiative losses, while the outer region defined by values close to zero. 

%For $Q$, it reproduce the pattern already pointed out in the offset map. The mean values for $Q$ and $v_{LOS}$ are indicated in Table~\ref{table:1}. 
If we neglect the central-upper part of the maps, outside the boundary of the plage region, the average amplitudes become $\overline{amp}_{Q}=7.6$ kW m$^{-2}$ and $\overline{amp}_{v_{LOS}}=3.2$ km~s$^{-1}$. The latter is in agreement with values reported by \citet{2009ApJ...692.1211C} in a facular region using the \ion{He}{i}~10839~\AA\ line. The relatively large value of the period in the chromosphere can be due to the propagation along an inclined magnetic field line which can extend the cut-off frequency in the chromosphere \citep[e.g.,][]{2007ApJ...671.1005B}.

The third column of Fig.~\ref{sin_map} shows the period maps of both $Q$ an $v_{LOS}$. The yellow contours indicate those areas where the radiative losses are particularly strong ($Q<-20$ kW~m$^{-2}$). The average period of oscillation for $Q$ is $\overline{p}_{Q}= 5.5$ min, while considering only the region inside the contour it drops to $\overline{p}_{Q}= 5.2$ min. These results found for $v_{LOS}$ are in line with previous works \citep{2004ESASP.547...25D,2009ApJ...692.1211C}, with a value of $\overline{p}_{v_{LOS}}= 5.5$ min. We note that in the plage area (bottom half of the panel) the periods present a quite smooth distribution, between 4 an 5 minutes, whereas the upper area of the image, outside the plage region, shows elongated features with longer periods. The latter are located at the plage boundary, where the magnetic field is more horizontal.
The phase difference between the radiative losses and the line-of-sight velocity is dominated by values close to $\pi/2$. Since the behaviour of the radiative losses is dominated by the temperature, we are essentially recovering the usual phase relation for running waves. %In a recent study, \citet{2021A&A...645L..12F} found that this phase difference in sunspots can change over time due to the formation of a resonant cavity when waves/shocks are reflected in the transition region.}

%\jcr{Summarizing, our interpretation of the results form the temporal analysis is that there is a spatially-smooth component in the net radiative losses that is largely correlated with the smooth and relatively stable magnetic field stratification. The mean offset value is -15.2~kW~m$^{-2}$. A weaker time-dependent and finely structured component is also present, and we relate it to the release of energy by waves and shocks with a mean amplitude of 4.5~kW~m$^{-2}$. }

Although we have not included plots showing the imprint of the periodic signal in each of the sublayers in which we divided the chromosphere, we have performed the fits individually for each layer. In the lower and middle chromosphere the modulation amplitude is of the order of 1~kW~m$^{-2}$, whereas in the upper chromosphere we get values much closer to the integral over the entire chromosphere. Therefore, we can conclude that although the imprint of waves is present in the lower and middle chromosphere, their contribution is larger in the upper chromosphere.

    \begin{table}
    \setlength\tabcolsep{5.5pt}
    \centering
     \begin{tabular}{c |c |c | c | c} 
     \hline\hline
     \centering
     & Offset  & Amplitude & Period & Phase  \\
            \hline
    $Q$ & -26.1 [kW$m^{-2}$] & 7.0 [kW$m^{-2}$] & 5.5 [min]& 3.3 [rad]\\
    $v_{LOS}$ & -0.1 [km/s] & 2.9 [km/s] & 5.5 [min] & 3.4 [rad]\\
     \hline
    \end{tabular}
    \caption{Average values of the parameters of the sinusoidal functions obtained over the red region of Figure~\ref{Q_contour}. Each value of the table corresponds to the average of the respective panel of Figure~\ref{sin_map}.}             
    \label{table:1}
    \end{table}
    
\section{Discussion and conclusions}

\begin{figure}
    \centering
    \includegraphics[width=\columnwidth]{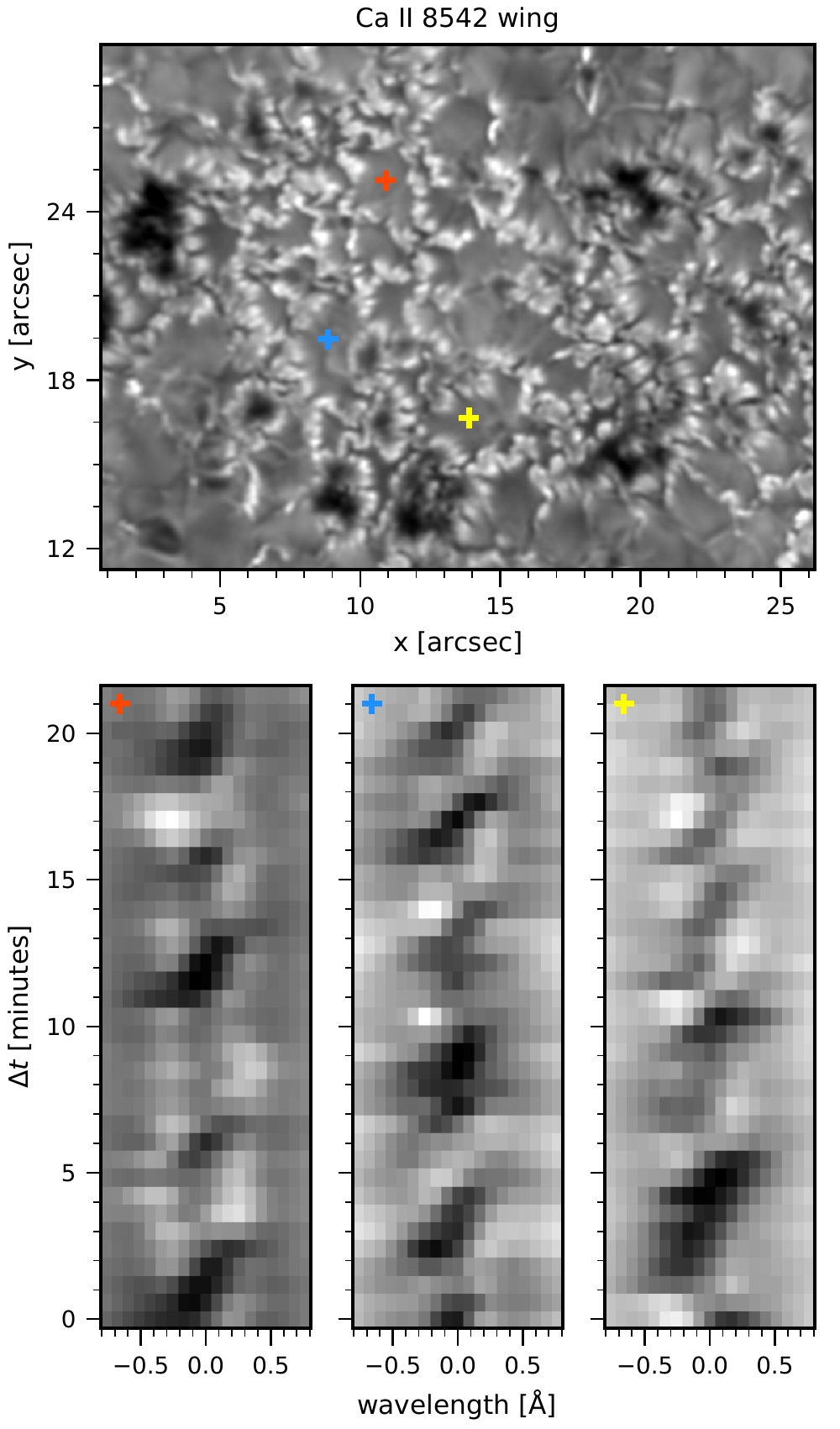}
    \caption{\emph{Lower row:} time evolution of the \ion{Ca}{ii}~8542~\AA\ line at three  locations, picked in the middle of the canopy regions, which are surrounded by the photospheric magnetic elements. The locations of the three pixels are marked in the upper panel with colored cross-like markers.}
    \label{fig:time}
\end{figure}

Our temporal analysis has allowed us to quantify the contribution from a periodic component, which we associate with wave heating. This component is weaker than the offset (background) value of the heating terms and it has a mean modulation amplitude of $\sim 7.0$~kW~m$^{-2}$. This component is responsible for the very fine structure that we observe in the $Q_{\mathrm{tot}}$ maps. The offset value, which we associate with a more static or very slowly evolving component, has a mean value of $\sim -26.1$~kW~m$^{-2}$. The map constructed from the offset value is relatively smooth, which could also point to a magnetic origin. On the Sun, the $\beta = 2\mu P_g / B^2=1$ layer is usually located in the lower chromosphere, and therefore, the magnetic field becomes smooth and room-filling above that layer. Having a relatively smooth offset map signals a magnetic origin. The amplitude of the periodic component of the radiative losses is almost a factor four larger in the upper chromosphere than in the lower and middle regions. Given the spatial distribution of periods and the relatively homogeneous $\pi/2$ phase difference, we associate the periodic component with compressible acoustic waves (the slow-mode of magnetoacoustic waves when $v_A > c_s$). This argument is further supported by Fig.~\ref{fig:time}, where we show the time-evolution of the 8542~\AA\ line at three random pixels selected in the middle of the canopy areas that are located in the surroundings of the photospheric magnetic elements. In all cases, the classical saw-tooth pattern from acoustic shocks is clearly visible.

Our results show that in the lower and middle chromosphere the radiative losses are distributed in areas surrounding the photospheric footpoints of the magnetic canopy. We do not see enhanced radiative losses within the magnetic footpoints of the canopy. Furthermore, \citet{2013ApJ...764L..11D} showed that in those regions the \ion{Ca}{ii}~8542~\AA\ lines profiles have a peculiar shape that can be explained by the presence of a hot magnetic canopy in the chromosphere that extends over a relatively quiet photosphere. The magnetic canopy should have a relatively sharp lower edge, where current sheets should be found through the relation $\boldsymbol{j} = \nabla \times \boldsymbol{B}/\mu$. Although, our inverted models have a low depth resolution in the magnetic field reconstruction due to the S/N ratio of our observations, the temperature stratification was derived with more than twice number of nodes and we do appreciate a relatively sharp canopy boundary there. Therefore, we argue that in the lower chromosphere of plage, Ohmic current dissipation must be responsible for the bulk of the heating. We also argue that if wave heating was a dominant phenomena in the lower chromosphere, its imprint should also be visible inside the magnetic footpoints, which we do not observe in our results \citep[see e.g.,][]{2008ApJ...680.1542H}. \citet{2019ApJ...879...57B} also showed that in Ohmic dissipation works efficiently in the photosphere / lower chromosphere.

In the upper chromosphere the radiative losses form a patch that covers the entire plage region, including most of the pores that are present in the photosphere. Within this patch, the derived chromospheric temperatures are relatively homogeneous and larger than in the surroundings, with temperatures the order of $\sim 7.5-8$~kK and a magnetic field strength of the order of $\sim 400$~G. The largest contribution to the integrated radiative losses is from the \ion{H}{i} diagnostics in the upper chromosphere. The contribution of wave heating is, in our results, largest in this layer. From our results alone, we cannot claim to have direct evidence of turbulent Alfv\'en wave heating \citep{2011ApJ...736....3V} or from neutral-ion collisions \citep{2012ApJ...747...87K}. Given the smooth nature of the magnetic field, we do not expect current sheets to be present in the upper chromosphere, making Ohmic dissipation of currents a less likely heating mechanism. The periodic behaviour that we observe has a relatively long period of 5.5~m. At a cadence of $\sim 30$~s, wave patterns with periods lower than 1~m are not properly sampled in our observations, so we cannot resolve high frequency waves. 

In order to investigate deeper the origin of the wave behaviour and the deposition of energy in the chromosphere, we briefly extended our study to to the microturbulence velocity $v_{turb}$. However, the temporal analysis, represented in the last row of Fig.~\ref{sin_map}, doesn't show a clear pattern in the offset or in the amplitude. It is definitely different from the typical white noise, but doesn't correlate with the patterns shown in the panels of $Q$ or $v_{LOS}$. We could expect an imprint of the wave pattern in the microturbulence if the shocks were relatively unresolved in depth by the inversion code, forcing a larger microturbulence value to account for the extra broadening. We do not observe such behaviour.

Our results are somewhat different than those reported by \citet{2021ApJ...921...39A}, but surprisingly also compatible. Their observations were based on arguably lower spatial-resolution slit-spectrograph raster scans in the \ion{Mg}{ii}~h\&k lines (IRIS) and in the \ion{He}{i}~10830~\AA\ line. Although their analysis was not based on the calculation of radiative losses, they used the integrated h\&k line intensity as a proxy, similarly done with the \ion{Ca}{ii}~H\&K lines by \citet{2018A&A...612A..28L}. We have shown that the radiative losses in the lower chromosphere are very small in the \ion{Mg}{ii} lines. By not having the \ion{Ca}{ii} deeper contribution, they would also miss the heating closer to the lower boundary of the magnetic canopy. As for their estimates of the chromospheric magnetic field, they are based on inversions of the \ion{He}{i}~10830~\AA\ line, and the latter usually samples the middle/upper chromosphere \citep[see Fig.~1 in][]{2019A&A...623A..74D} according to estimates from numerical simulations.

The present study is unique in that we have calculated and studied the distribution of radiative losses as a function of depth and time from very high spatial-resolution spectra. The canonical value of $Q_{\mathrm{AR}}\sim 20$~kW~m$^{-2}$, derived in the 70's and 80's using spatially and temporally averaged spectra, could not catch the complexity and dynamic behaviour of the solar chromosphere as shown in our analysis. Our analysis is also different from previous studies in that we have estimated the contribution from waves directly from the periodic modulation of the radiative losses, and not by estimating from Doppler velocities the energy that is carried out by waves in the chromosphere \citep{2020A&A...642A..52A}.

Ultimately, our results could not provide direct evidence that would allow discriminating what heating mechanism is dominating in the upper chromosphere. In our opinion, future studies of the same nature should analyze higher cadence time-series and higher spatial resolution observations in order to attempt finding observational signatures of high frequency waves or very small-spatial-scale variations that could point to turbulent Alfv\'en wave dissipation. In order to estimate ambipolar diffusion heating, an accurate estimate of the upper chromosphere density is also needed, and therefore inversion codes must be modified in order to include the Lorentz force \citep{2019A&A...629A..24P} and the support effects derived from velocity gradients.

\begin{acknowledgements}
%JdlCR
This project has received funding from the European Research Council (ERC) under the European Union's Horizon 2020 research and innovation program (SUNMAG, grant agreement 759548). %JdlCR is supported by grants from the Swedish Research Council (2015-03994) and the Swedish National Space Agency (128/15). 
JL is supported through the CHROMATIC project (2016.0019) funded by the Knut och Alice Wallenberg foundation.
%SST
The Swedish 1-m Solar Telescope is operated on the island of La Palma by the Institute for Solar Physics of Stockholm University in the Spanish Observatorio del Roque de los Muchachos of the Instituto de Astrof\'isica de Canarias. The Institute for Solar Physics is supported by a grant for research infrastructures of national importance from the Swedish Research Council (registration number 2017-00625).
This research has made use of NASA's Astrophysics Data System Bibliographic Services.
%SNIC
The computations were enabled by resources provided by the Swedish 
National Infrastructure for Computing (SNIC) at the PDC Center for High 
Performance Computing, KTH Royal Institute of Technology, partially 
funded by the Swedish Research Council through grant agreement no. 
2018-05973.
\end{acknowledgements}


\bibliographystyle{aa} % style aa.bst
\bibliography{references}

\begin{appendix}
%\section{Appendix?}\label{ap:1}
\section{Training of the neural network}\label{ap:nn}

\begin{figure}
    \centering
    \includegraphics[width=0.9\columnwidth,trim=13cm 9cm 17.5cm 0.5cm, clip]{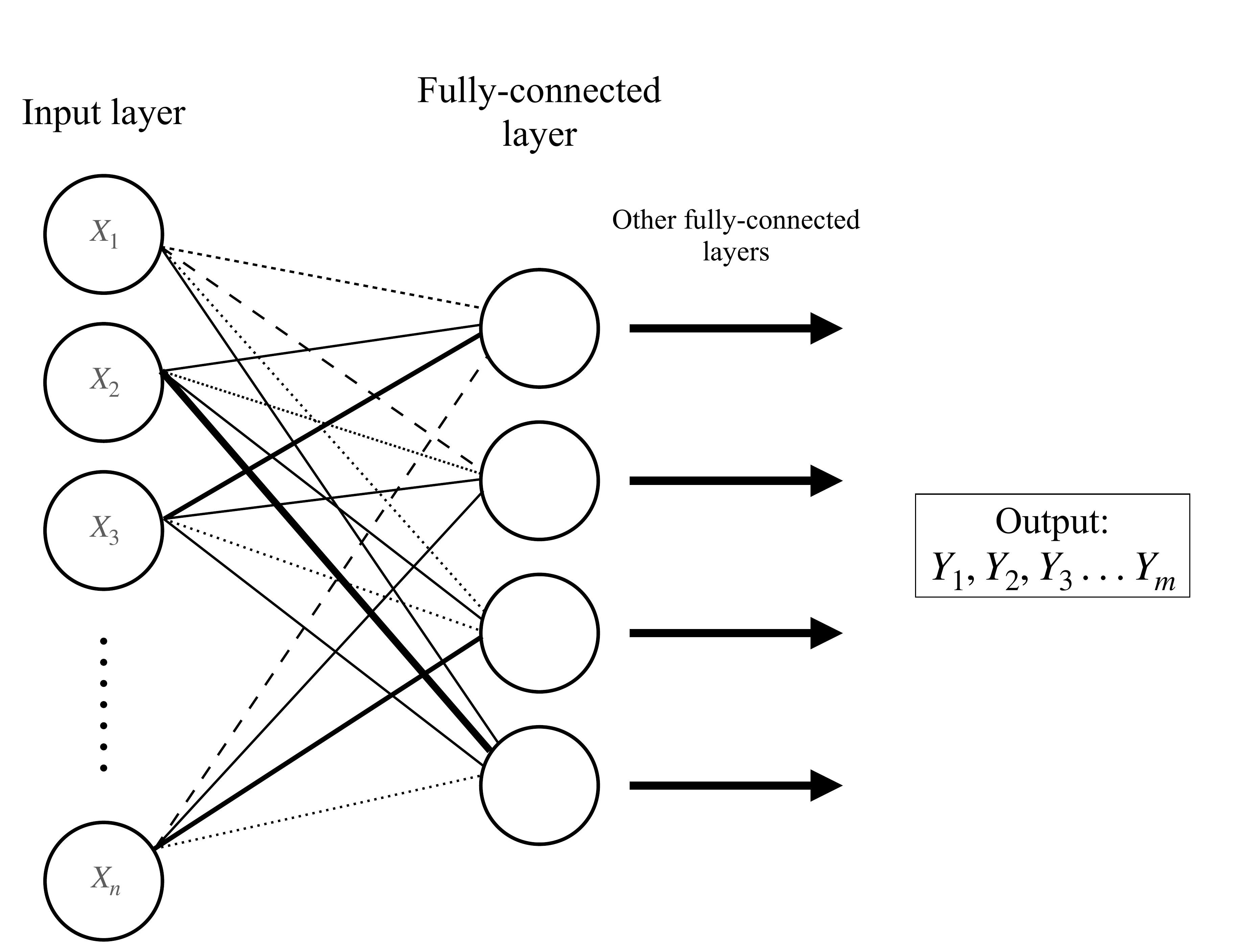} %trim=2.5cm 1cm 2.5cm 0.5cm, clip
        \caption{Simplified representation of a fully-connected neural network. The lines that connect the inputs with the neurons are represented in different style (straight, doted, dashed, ...) because they involve different weights.}
    \label{nn_architecture}         
\end{figure}

In general, an artificial neural network is defined by the dimensionality of the input, the number of layers, the number of neurons per layer and the dimensionality of the output. The number of neurons in each layer does not have to be constant, and can vary depending on the complexity of the problem. The most used type of neural network is the fully connected network (FCN; \citealt{Schmidhuber_2015}), in which every input is connected to every neuron of the following layer. Figure~\ref{nn_architecture} shows, in a simplified way, the architecture and connections of a FCN. Each connection is described by a simple function that combines linearly the input $x$ multiplied by a weight $w$ and summed with a bias $b$ and finally returns the value of a certain user-defined nonlinear function $f(x)$. In mathematical notation, the information that will pass from the input neurons $i$ to the neuron $j$ of the next layer will be:
\begin{equation}
    \sigma_j = f(\Sigma_i (w_i x_i + b_j)).    
\end{equation}
This output will be the input for another neuron of the next layer. Since the first operation is linear, the activation is the one that introduces the non-linear character of the FCNs.

The optimization of a neural network is called training and it involves the iterative modification of the weights and biases so that a loss function that measures the ability of the network to predict the output from the input is minimized. In our case, we have trained the neural network to learn the mapping between the observed Stokes profiles and model atmosphere obtained from the inversion. Once the network is trained, we are able to reconstruct the temperature $T$, the line-of-sight velocity $v_{LOS}$, the turbulence velocity $v_{turb}$, the parallel component of the magnetic field $B_{||}$, the perpendicular one $B_{\perp}$ and the azimuth angle $\phi$ for the entire time series. The dimensions of the input are defined by the total number of pixels and by the four Stokes parameters. On the other hand, the output dimensions are the total number of pixels and the number of obtained parameters multiplied by the number of grid points.

\paragraph{Architecture of the NN:} For our purposes, we design a fully-connected neural network with 5 layers and 200 neurons per layer. We have found this configuration to be optimal after many tests in terms of training time and accuracy. The activation function that we decided to use is the rectified linear unit or ReLU \citep{Nair2010RectifiedLU}. It has a linear behavior for a positive input, otherwise, if the input is negative, it is equal to zero:
\begin{equation}
    \textrm{ReLU}(x) = \max (0, x). 
\end{equation}
It is applied after every layer of the NN, except for the last one, to avoid obtaining only positive outputs.

During the design of the network architecture we detected that the noise in the polarization  was amplified and propagated to some physical parameters. To avoid this problem we decided to split the model into two parts: only Stokes $I$ was employed in the calculation of the temperature and $v_{LOS}$, while all the four Stokes parameters are used for the other quantities of the model atmosphere. Although Stokes $Q$, $U$ and $V$ have information of the gradient of the source function and the line-of-sight velocity stratification, in cases where the profiles are very noisy they do not play an important role in the derivation of the temperature and line-of-sight velocity, so it is better to avoid propagating that noise.

\paragraph{Training process and validation set:} Optimization is routinely solved using simple first-order gradient descent algorithms, which modify the weights along the negative gradient of the loss function with respect to the model parameters. To scale the magnitude of our weight updates we have to use the parameter called learning rate. This parameter has to be adjusted to find a compromise between network accuracy and convergence speed. If this number is too small, it will take too long to reach the solution, while, if it is too large, there is a risk of overshooting the optimal solution. In our case we used a learning rate of $10^{-4}$. For the optimization method, we have used a gradient descent variant called Adam \citep{kingma2017adam} which has been developed to automatically adjust the learning rate, making the solution convergence faster.

Our goal is to optimize a user-defined loss function that evaluates how well our network models the data. The most common loss function is the mean squared error which measures the average squared difference between predictions and desirable outputs. However, to get an idea of the dispersion of the estimate we have used the quantile regression \citep{Koenker1978}, a loss function for estimating any  percentile value:
\begin{equation}
    L(x,y|q) = \left\{
	       \begin{array}{ll}
		 q (y-f(x))      & \mathrm{if\ } (y-f(x)) \ge 0 \\
		 (q -1) (y-f(x)) & \mathrm{if\ } (y-f(x)) < 0 \\
	       \end{array}
	     \right.
\end{equation}
where $y$ is the training value and $f(x)$ is output value of the network. During training, $q$ is randomly varied between 0 and 1 so that the network can learn all possible percentiles. This allows us not only to estimate the mean value ($q$=0.5) but also the dispersion ($q$=0.16 and $q$=0.84) which is equivalent to one standard deviation $\sigma$ in the case of a normal distribution. The dispersion can give us an idea of the uncertainty of the inversion since similar Stokes parameters could have had different atmospheric models as solutions \citep{2021arXiv210807089D}.

The gradient of the loss function with respect to the free parameters of the network is obtained using the backpropagation algorithm. Since the networks are defined as a stack of layers, the gradient of the loss function can be calculated by the chain rule as the product of the gradient of each module and ultimately of the last layer and the specific loss function. The main problem with some activation functions is that the gradient vanishes for very large values due to the derivative of this function, making it difficult to train the network. For this reason, we have used the ReLU function, which does not saturate for large values.

% \begin{figure}[ht!]
%     \centering
%     \includegraphics[width=\columnwidth]{loss_vs_lossval.pdf}
%     \includegraphics[width=\columnwidth]{loss_list_final_NLTE.pdf}%trim=2.5cm 1cm 2.5cm 0.5cm, clip
%         \caption{Example of two different training of the NN. \emph{Upper panel:} The loss function of the validation set is not following the loss function of the training set, meaning that the NN is not learning anymore, but just memorizing. \emph{Lower panel:} The loss function of the validation set and of the training set follow the same slope, meaning that the NN is still learning.}
%     \label{loss_list}         
% \end{figure}

Regarding the dataset, we used two non-consecutive time steps of the observations to train the neural network, in order to include more statistics in our training set, which correspond to about 41160 pixels. To further increase the diversity of profiles used for training the network, we added a Gaussian noise component to the input profiles during the training to make the network prediction more robust to the noise.

Because of the large number of free parameters in a network, overfitting can be a problem. One would like the network to generalize well and avoid any memorization of the training set \citep{10.5555/525960,ripley_1996}. To check that, a part of the dataset is not used during the training but used after each iteration as validation. Desirably, the loss should decrease both in the training and validation sets simultaneously. If overfitting occurs, the loss in the validation set will increase. We have randomly chosen 90\% of the dataset as training set and the 10\% as validation set. Furthermore, every time that the loss function, calculated with the validation set, reached its minimum, we saved the model parameters. The training was done in a GeForce RTX 2080 Ti GPU for 400000 epochs. %, taking around XXX seconds the whole process. 
Once we had picked the best weights and biases, we are able to apply the obtained model to the other time step of the observations.

\end{appendix}

\end{document}